# Chapter 27
# Aneka Cloud Application Platform and Its Integration with Windows Azure

Yi Wei[1], Karthik Sukumar[1], Christian Vecchiola[2], Dileban Karunamoorthy[2] and Rajkumar Buyya[1, 2]

[1]Manjrasoft Pty. Ltd., Melbourne, Victoria, Australia

[2]Cloud Computing and Distributed Systems (CLOUDS) Laboratory, Department of Computer Science and Software Engineering, The University of Melbourne, Australia

## Abstract

Aneka is an Application Platform-as-a-Service (Aneka PaaS) for Cloud Computing. It acts as a framework for building customized applications and deploying them on either public or private Clouds. One of the key features of Aneka is its support for provisioning resources on different public Cloud providers such as Amazon EC2, Windows Azure and GoGrid. In this chapter, we will present Aneka platform and its integration with one of the public Cloud infrastructures, Windows Azure, which enables the usage of Windows Azure Compute Service as a resource provider of Aneka PaaS. The integration of the two platforms will allow users to leverage the power of Windows Azure Platform for Aneka Cloud Computing, employing a large number of compute instances to run their applications in parallel. Furthermore, customers of the Windows Azure platform can benefit from the integration with Aneka PaaS by embracing the advanced features of Aneka in terms of multiple programming models, scheduling and management services, application execution services, accounting and pricing services and dynamic provisioning services. Finally, in addition to the Windows Azure Platform we will illustrate in this chapter the integration of Aneka PaaS with other public Cloud platforms such as Amazon EC2 and GoGrid, and virtual machine management platforms such as Xen Server. The new support of provisioning resources on Windows Azure once again proves the adaptability, extensibility and flexibility of Aneka.

## Keyword

## 1. INTRODUCTION

Current industries have seen Clouds [2, 14] as an economic incentive for expanding their IT infrastructure with less total cost of ownership (TCO) and higher return of investment (ROI). By supporting virtualization and dynamic provisioning of resources on demand, Cloud computing paradigm allows any business, from small and medium enterprise (SMEs) to large organizations, to more wisely and securely plan their IT expenditures. They will be able to respond rapidly to variations in the market demand for their Cloud services. IT cost savings are realized by means of the provision of IT "subscription-oriented" infrastructure and services on a pay-as-you-go-basis. There is no more need to invest in redundant and highly fault tolerant hardware or expensive software systems, which will lose their value before they will be paid off by the generated revenue. Cloud computing now allows paying for what the business need at the present time and to release it when these resources are no longer needed. The practice of renting IT infrastructures and services has become so appealing that it is not only leveraged to integrate additional resources and elastically scale existing software systems into hybrid Clouds, but also to redesign the existing IT infrastructure in order to optimize the usage of the internal IT, thus leading to the birth of private Clouds. To effectively and efficiently harness Cloud computing, service providers and application developers need to deal with several challenges, which include: application programming models, resource management and monitoring, cost-aware provisioning, application scheduling, and energy efficient resource utilization. The Aneka Cloud Application platform, together with other virtualization and Cloud computing technologies aims to address these challenges and to simplify the design and deployment of Cloud Computing systems.

Aneka is a .NET-based application development Platform-as–a-Service (PaaS), which offers a runtime environment and a set of APIs that enable developers to build customized applications by using multiple programming models such as *Task Programming*, *Thread Programming* and *MapReduce Programming*, which can leverage the compute resources on either public or private Clouds [1]. Moreover, Aneka provides a number of services that allow users to control, auto-scale, reserve, monitor and bill users for the resources used by their applications. One of key characteristics of Aneka PaaS is to support provisioning of resources on public Clouds such as Windows Azure, Amazon EC2, and GoGrid, while also harnessing private Cloud resources ranging from desktops and clusters, to virtual datacentres when needed to boost the performance of applications, as shown in **Figure 1**. Aneka has successfully been used in several industry segments and application scenarios to meet their rapidly growing computing demands.

In this chapter, we will introduce Aneka Cloud Application Platform (Aneka PaaS) and describe its integration with public Cloud platforms particularly focusing on the Windows Azure Platform. We will show in detail, how an adaptable, extensible and flexible Cloud platform can help enhance the performance and efficiency of applications by harnessing resources from private, public or hybrid Clouds with minimal programming effort. The Windows Azure Platform is a Cloud Services Platform offered by Microsoft [5]. Our goal is to integrate the Aneka PaaS with Windows Azure Platform, so that Aneka PaaS can leverage the computing resources offered by Windows Azure Platform. The integration supports two types of deployments. In the first case, our objective is to deploy Aneka Worker Containers as instances of Windows Azure Worker Role, while the Aneka Master Container runs locally on-premises, enabling users of Aneka PaaS to use the computing resources offered by Windows Azure Platform for application execution. And in the second case, the entire Aneka Cloud is deployed on Windows Azure so that Aneka users do not have to build or provision any computing resources to run Aneka PaaS. This chapter reports the design and implementation of the deployment of Aneka containers on Windows Azure Worker Role and the integration of two platforms.

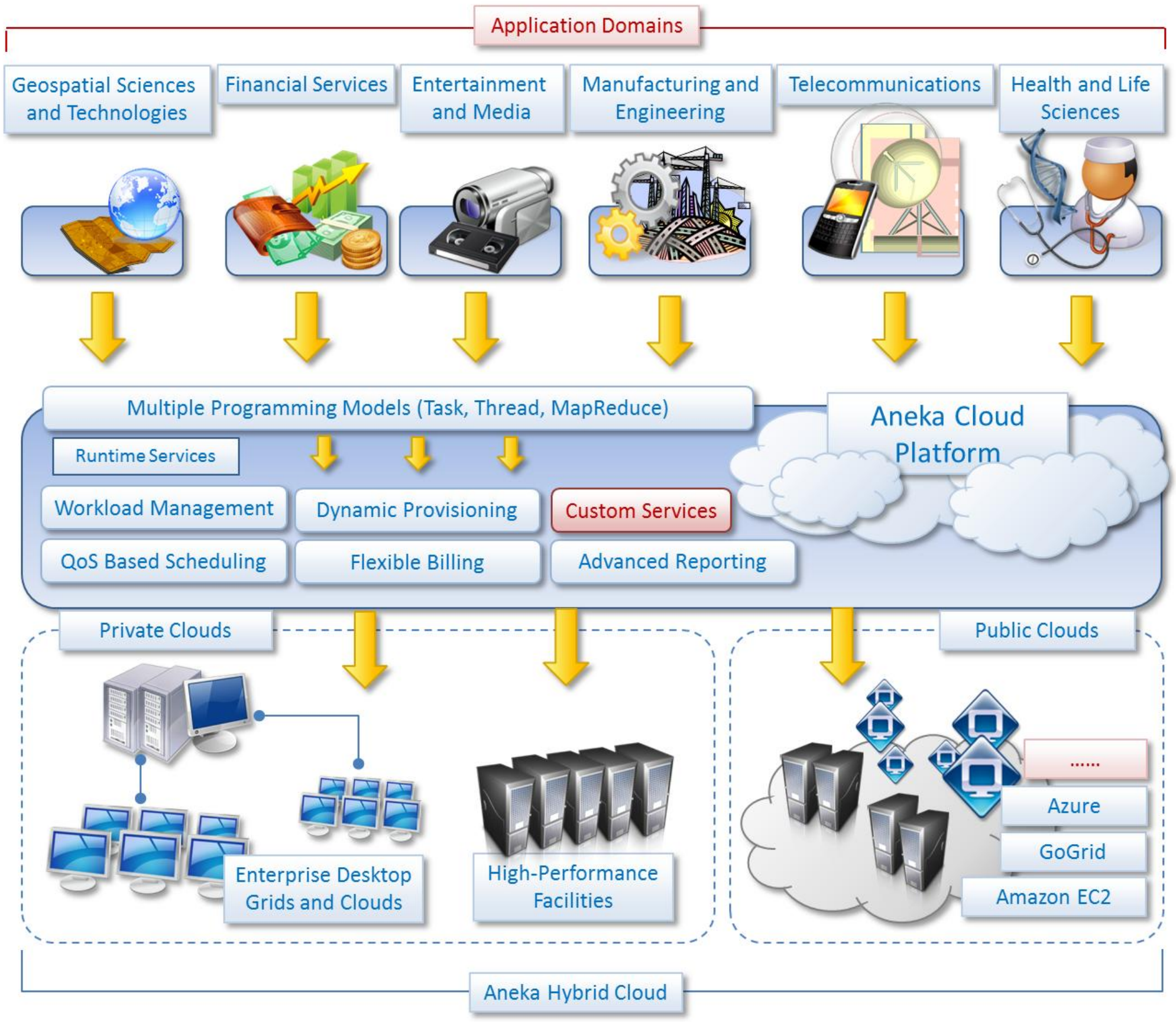

***Figure 1: Aneka Cloud Application Development Platform.***

The remainder of the chapter is structured as follows: in section 2, we present the architecture of Aneka PaaS, provide an overview of the Windows Azure Platform and Windows Azure Service Architecture, and list the advantages of integrating the two platforms along with the limitations and challenges we faced. Section 3 demonstrates our design in detail on how to integrate the Aneka PaaS with Windows Azure Platform. Next, we will discuss the implementation of the design in Section 4. Section 5 presents the experimental results of executing applications on the two integrated environments. In Section 6 and 7, we list related work and sample applications of Aneka. Finally, we present the conclusions and future directions.

## 2. BACKGROUND

In this section, we present the architecture of Aneka PaaS, and then depict the overall view on Windows Azure Platform and Windows Azure Service Architecture. We also discuss the advantages brought by the integration, along with the limitations and challenges faced.

## 2.1 Overview of Aneka Cloud Application Development Platform

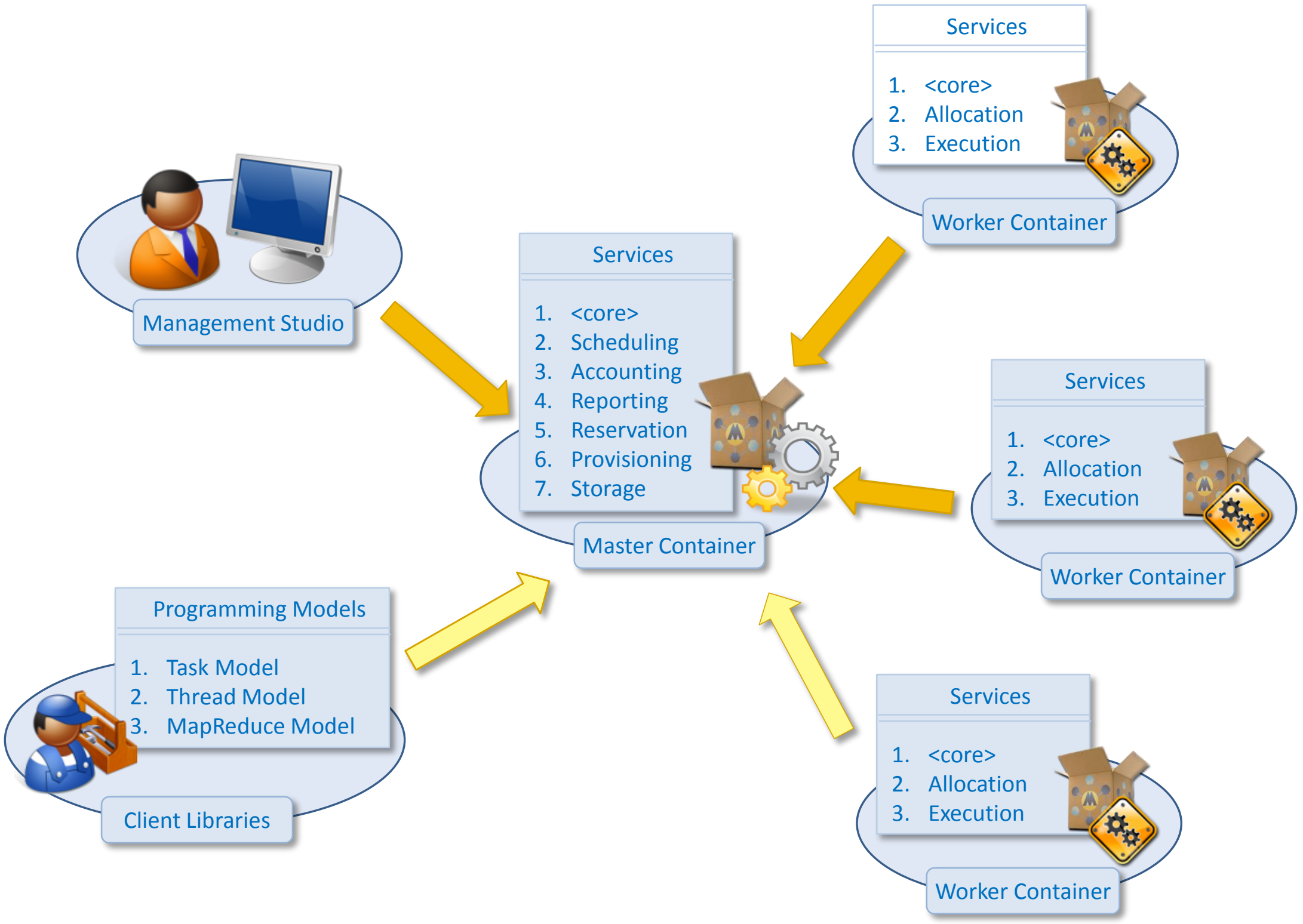

**Figure** 2 shows the basic architecture of Aneka. The system includes four key components, including Aneka Master, Aneka Worker, Aneka Management Console, and Aneka Client Libraries [1].

The Aneka Master and Aneka Worker are both Aneka Containers which represents the basic deployment unit of Aneka based Clouds. Aneka Containers host different kinds of services depending on their role. For instance, in addition to mandatory services, the Master runs the Scheduling, Accounting, Reporting, Reservation, Provisioning, and Storage services, while the Workers run execution services. For scalability reasons, some of these services can be hosted on separate Containers with different roles. For example, it is ideal to deploy a Storage Container for hosting the Storage service, which is responsible for managing the storage and transfer of files within the Aneka Cloud. The Master Container is responsible for managing the entire Aneka Cloud, coordinating the execution of applications by dispatching the collection of work units to the compute nodes, whilst the Worker Container is in charge of executing the work units, monitoring the execution, and collecting and forwarding the results.

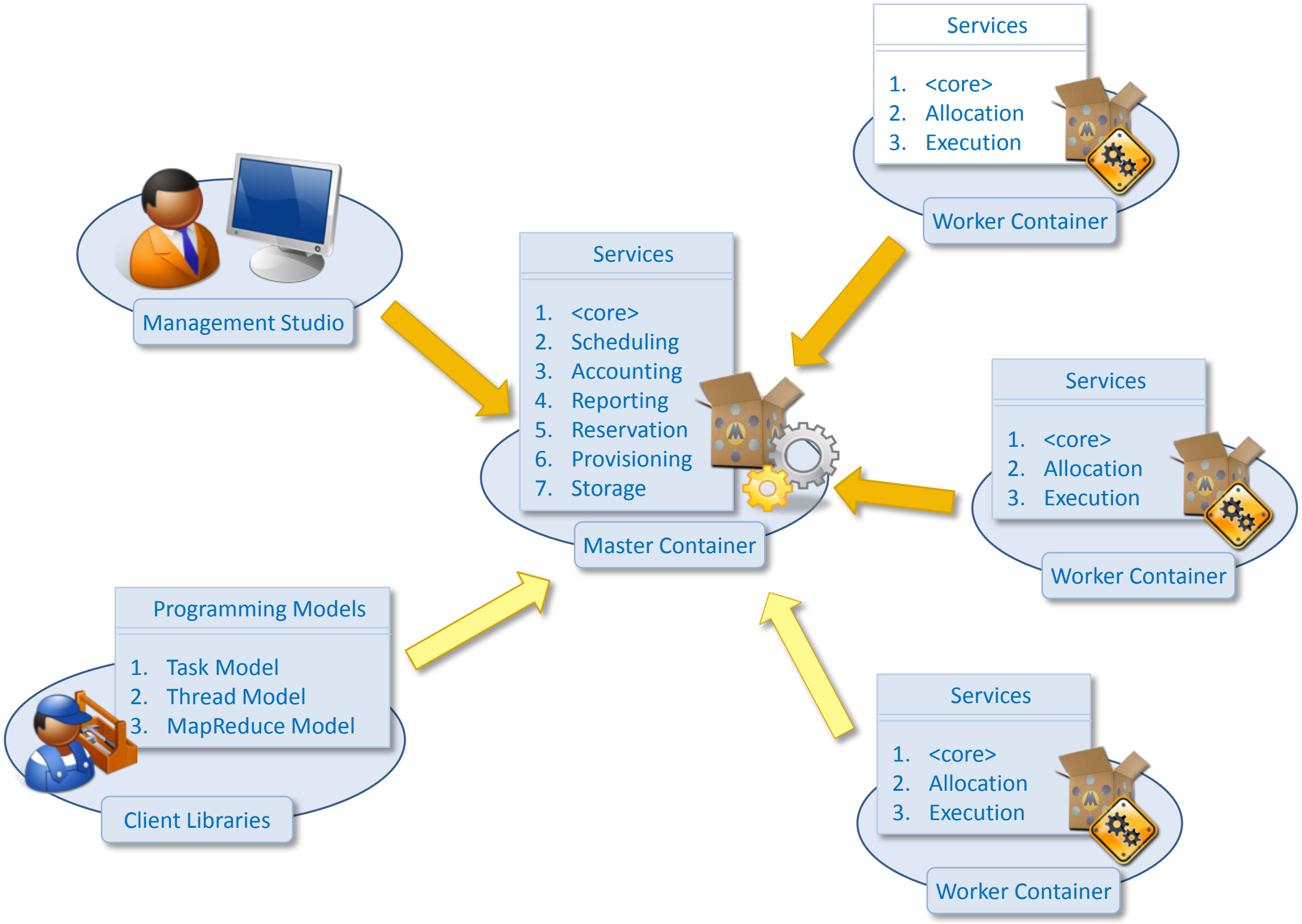

***Figure 2: Basic Architecture of Aneka.***

The Management Studio and client libraries help in managing the Aneka Cloud and developing applications that utilize resources on Aneka Cloud. The Management Studio is an administrative console that is used to configure Aneka Clouds; install, start or stop Containers; setup user accounts and permissions for accessing Cloud resources; and access monitoring and billing information. The Aneka client libraries, are Application Programming Interfaces (APIs) used to develop applications which can be executed on the Aneka Cloud. Three different kinds of Cloud programming models are available for the Aneka PaaS to cover different application scenarios:: *Task Programming*, *Thread Programming* and *MapReduce Programming* These models represent common abstractions in distributed and parallel computing and provide developers with familiar abstractions to design and implement applications.

### 2.1.1 Fast and Simple: Task Programming Model

Task Programming Model provides developers with the ability of expressing applications as a collection of independent tasks. Each task can perform different operations, or the same operation on different data, and can be executed in any order by the runtime environment. This is a scenario in which many scientific applications fit in and a very popular model for Grid Computing. Also, Task programming allows the parallelization of legacy applications on the Cloud.

### 2.1.2 Concurrent Applications: Thread Programming Model

Thread Programming Model offers developers the capability of running multithreaded applications on the Aneka Cloud. The main abstraction of this model is the concept of *thread* which mimics the semantics of the common local thread but is executed remotely in a distributed environment. This model offers finer control on the execution of the individual components (threads) of an application but requires more management when compared to Task Programming, which is based on a "*submit and* *forget"* pattern. The Aneka Thread supports almost all of the operations available for traditional local threads. More specifically an Aneka thread has been designed to mirror the interface of the *System.Threading.Thread* .NET class, so that developers can easily move existing multi-threaded applications to the Aneka platform with minimal changes. Ideally, applications can be transparently ported to Aneka just by substituting local threads with Aneka Threads and introducing minimal changes to the code. This model covers all the application scenarios of the Task Programming and solves the additional challenges of providing a distributed runtime environment for local multi-threaded applications.

### 2.1.3 Data Intensive Applications: MapReduce Programing Model

MapReduce Programming Model [11] is an implementation of the MapReduce model proposed by Google [12], in .NET on the Aneka platform. MapReduce has been designed to process huge quantities of data by using simple operations that extracts useful information from a dataset (the *map* function) and aggregates this information together (the *reduce* function) to produce the final results. Developers provide the logic for these two operations and the dataset, and Aneka will do the rest, making the results accessible when the application is completed.

## 2.2 Overview of Windows Azure Platform

Generally speaking, Windows Azure Platform is a Cloud platform which provides a wide range of Internet Services [3]. Currently, it involves four components (**Figure 3**). They are Windows Azure, SQL Azure, Windows Azure AppFabric, and Windows Azure Market Place respectively.

Windows Azure, which we will introduce in detail in **Section 2.3**, is a Windows based Cloud services operating system providing users with on-demand compute service for running applications, and storage services for storing data in Microsoft data centres.

The second component, SQL Azure offers a SQL Server environment in the Cloud, whose features includes supporting Transact-SQL and support for the synchronization of relational data across SQL Azure and on-premises SQL Server.

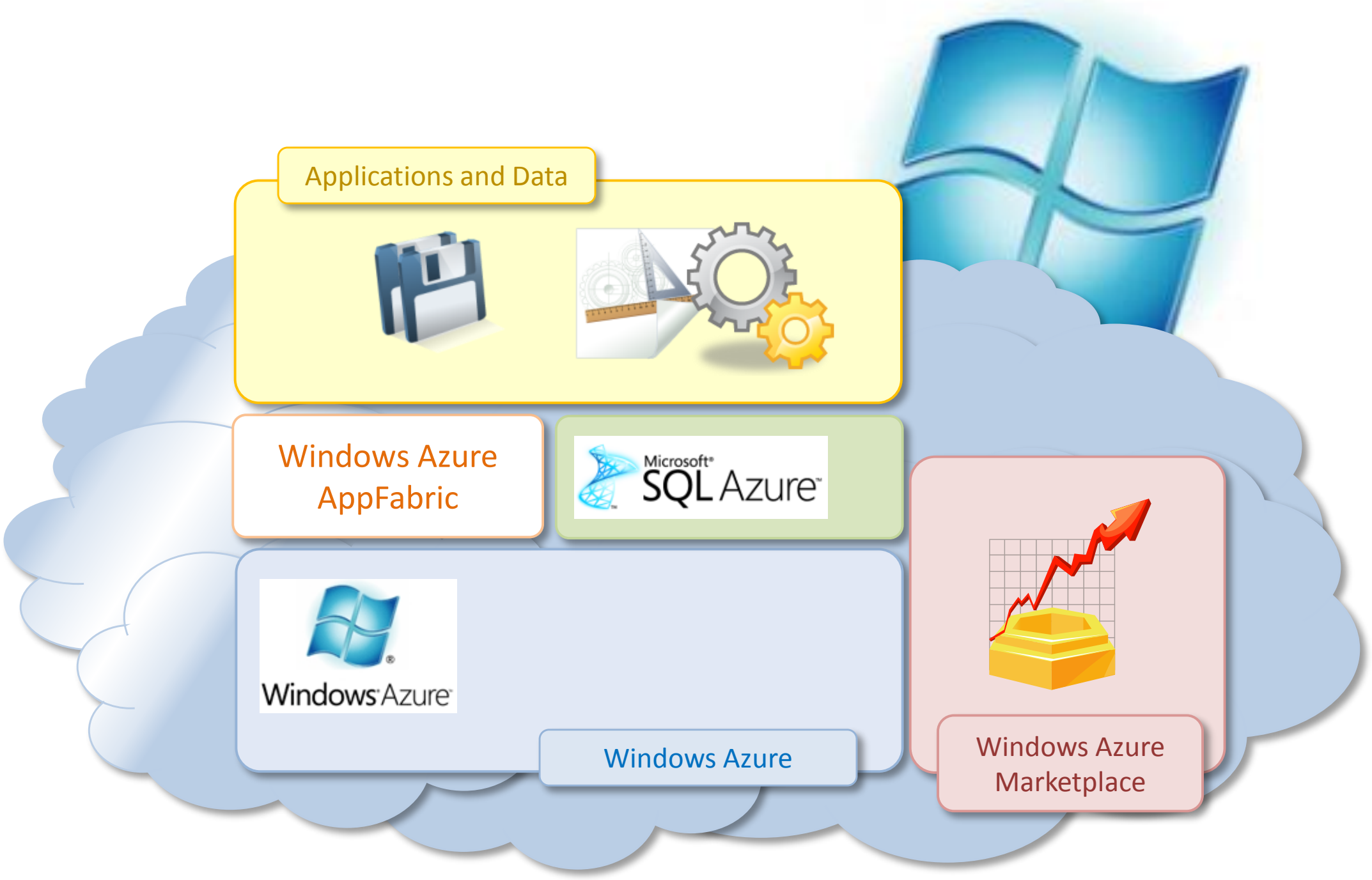

*Figure 3: The Components of Windows Azure Platform.*

Windows Azure AppFabric is a Cloud-based infrastructure for connecting Cloud and on-premise applications, which are accessed through HTTP REST API.

The newly born Windows Azure Marketplace is an online service for making transactions on Cloud-based data and Windows Azure Applications.

## 2.3 Overview of Windows Azure Service Architecture

In contrast to other public Cloud platforms such as Amazon EC2 and GoGrid, Windows Azure currently does not provide an IaaS (Infrastructure-as-a-Service). Instead, it provides a PaaS (Platform as a Service) solution, restricting users from direct access with administrative privileges to underlying virtual infrastructure. Users can only use the Web APIs exposed by Windows Azure to configure and use Windows Azure services.

A role on Windows Azure refers to a discrete scalable component built with managed code. Windows Azure currently supports three kinds of roles [4], as shown in **Figure 4**.

- Web Role: a Web role is a role that is customized for Web application programming as is supported by IIS 7.
- Worker Role: a worker role is a role that is useful for generalized development. It is designed to run a variety of Windows-based code.
- VM Role: a virtual machine role is a role that runs a user-provided Windows Server 2008 R2 image.

A Windows Azure service must include at least one role of either type, but may consist of any number of Web roles, worker roles and VM roles. Furthermore, we can launch any number of instances of a particular role. Each instance will be run in an independent VM and share the same binary code and configuration file of the role.

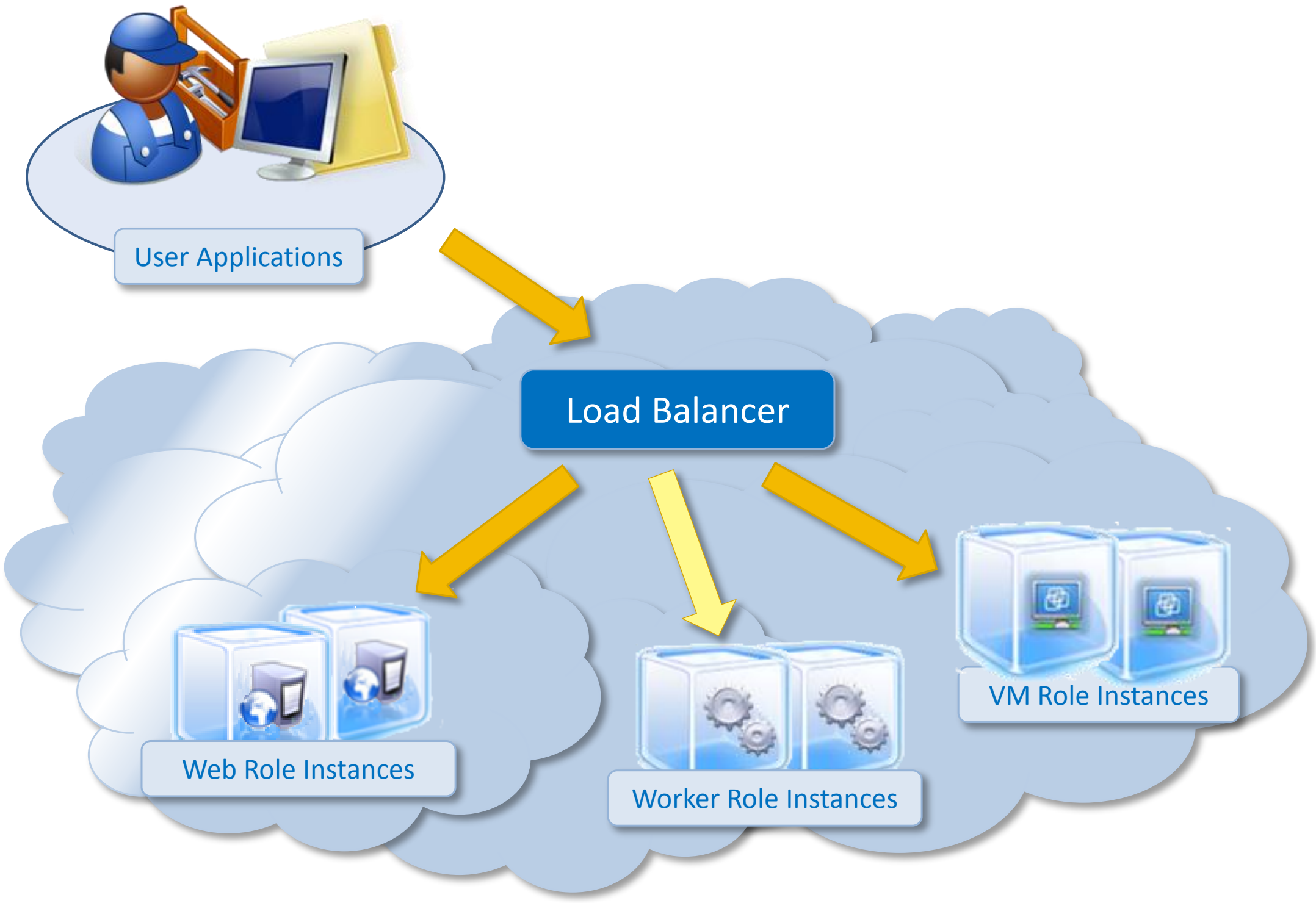

***Figure 4: Windows Azure Service Architecture.***

In terms of the communication support, there are two types of endpoints that can be defined: input and internal. Input endpoints are those are exposed to the Internet, and internal endpoints are used for communication inside the application within the Azure environment. A Web role can define a single HTTP endpoint and a single HTTPS endpoint for external users, whilst a Worker Role and a VM role may assign up to five internal or external endpoints using HTTP, HTTPS or TCP. There exists a built-in load balancer on top of each external endpoint which is used to spread incoming requests across the instances of the given role. Besides, all the role instances can make outbound connections to Internet resources via HTTP, HTTPS or TCP.

Under this circumstance, we can deploy Aneka Container as instances of Windows Azure Worker Role which gets access to resources on the Windows Azure environment via the *Windows Azure Managed Library*.

## 2.4 Advantages of Integration of two platforms

Inevitably, the integrated Cloud environment will combine features from the two platforms together, enabling the users to leverage the best of both platforms such as access to cheap resources, easy programming, and management of Cloud computing services.

### 2.4.1 Features from Windows Azure

For the users of Aneka Platform, the integration of the Aneka PaaS and Windows Azure resources means they do not have to build or provision the infrastructure needed for Aneka Cloud. They can launch any number of instances on Windows Azure Cloud Platform to run their application in parallel to gain more efficiency.

### 2.4.2 Features from Aneka Cloud Application Development Platform

For the users of Windows Azure Application, the integration of Aneka PaaS and Windows Azure Platform allows them to embrace the advanced features from Aneka PaaS:

- **Multiple Programming Models.** As discussed in **Section 2.1**, the Aneka PaaS provides users with three different kinds of cloud programming models, which involves *Task Programming*, *Thread Programming*, and *MapReduce Programming* to cover different application scenarios, dramatically decreasing the time needed in developing Cloud-aware applications, as shown in **Figure 5**.

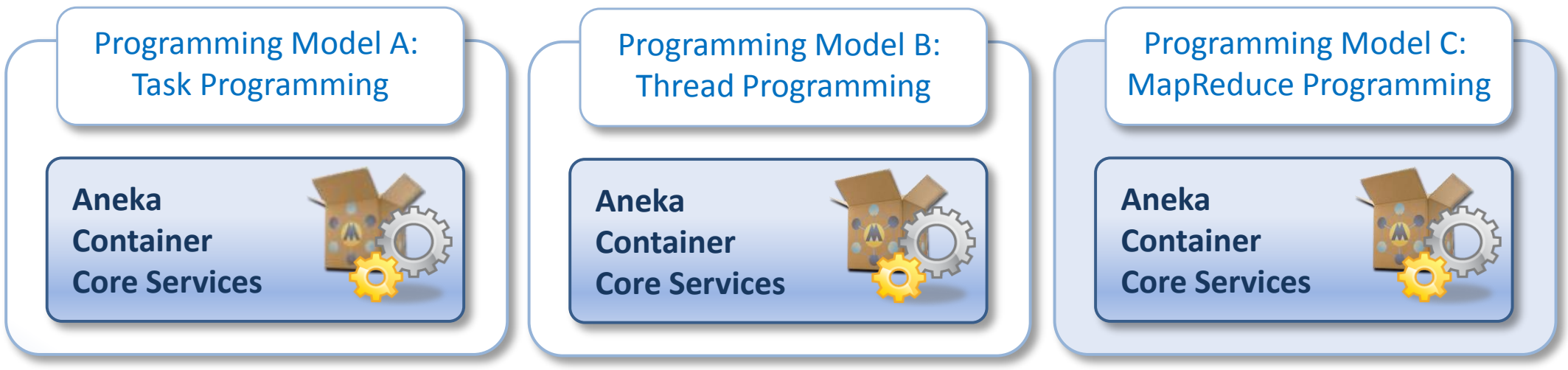

***Figure 5: Multiple Programming Models of the Aneka PaaS Patent.***

- **Scheduling and Management Services.** The Aneka PaaS Scheduling Service can dispatch the collection of jobs that compose an Aneka Application to the compute nodes in a completely transparent manner. The users do not need to take care of the scheduling and the management of the application execution.
- **Execution Services.** The Aneka PaaS Execution Services can perform the execution of distributed application and collect the results on the Aneka Worker Container runtime environment.
- **Accounting and Pricing Services**. Accounting and Pricing services of Aneka PaaS enable billing the final customer for using the Cloud by keeping track of the application running and providing flexible pricing strategies that are of benefit to both the final users of the application and the service providers.
- **Dynamic Provisioning Services.** In current pricing model for Windows Azure, customers will be charged at an hourly rate depending on the size of the compute instance. Thus it makes sense to dynamically add instances to a deployment at runtime according to the load and requirement of the application. Similarly instances can be dynamically decreased or the entire deployment can be deleted when not being actively used to avoid charges. One of the key features of Aneka is its support for dynamic provisioning which can be used to leverage resources dynamically for scaling up and down Aneka Clouds, controlling the lifetime of virtual nodes.

## 2.5 Limitations for the Integration

Although the integration of two platforms will generate numerous benefits for both Aneka users and Windows Azure customers, running Aneka Container as instances of Windows Azure Worker Role has some limitations.

The current version of Windows Azure does not provide administrative privileges on Windows Azure Worker Role instances. Deployments are prepared using the *Windows Azure Managed Library*, and the prepared Windows Azure Service Package is uploaded and run.

Under these circumstances, we cannot use the Aneka Management Studio to install Aneka Daemons and Aneka Containers on Windows Azure VMs directly. Further, other third party software that is needed on the Worker nodes such as PovRay and Maya, cannot be run on Windows Azure Worker Role instances because of the need for administrative privileges. This limits the task execution services that Azure Aneka Worker offers to XCopy deployment applications.

## 2.6 Challenges for the Integration

Due to the access limitations and service architecture of Windows Azure, we encountered some implementation issues that required changes to some parts of the design and implementation of the Aneka framework.

### 2.6.1 Administration Privileges

The Azure applications in both Web role and worker role instances do not have administrative privileges and does not have write access to files under the *"E:\approot\"* where the application code is deployed. On possible solution is to use *LocalResource* to define and use the local resource of Windows Azure VM disk.

Technically speaking, we need to dynamically change the path of files which are to be written to the local file system, to the path under the *RootPath* Property returned by the *LocalResource* object at runtime.

### 2.6.2 Routing in Windows Azure

Each Windows Azure Worker Role can define up to five input endpoints using HTTP, HTTPS or TCP, each of which is used as external endpoints to listen on a unique port.

One of the several benefits of using Windows Azure is that all the requests connected to an input endpoint of a Windows Azure Role will be connected to a load balancer which automatically forwards the requests to all instances that are declared to be part of the role, on a round robin basis.

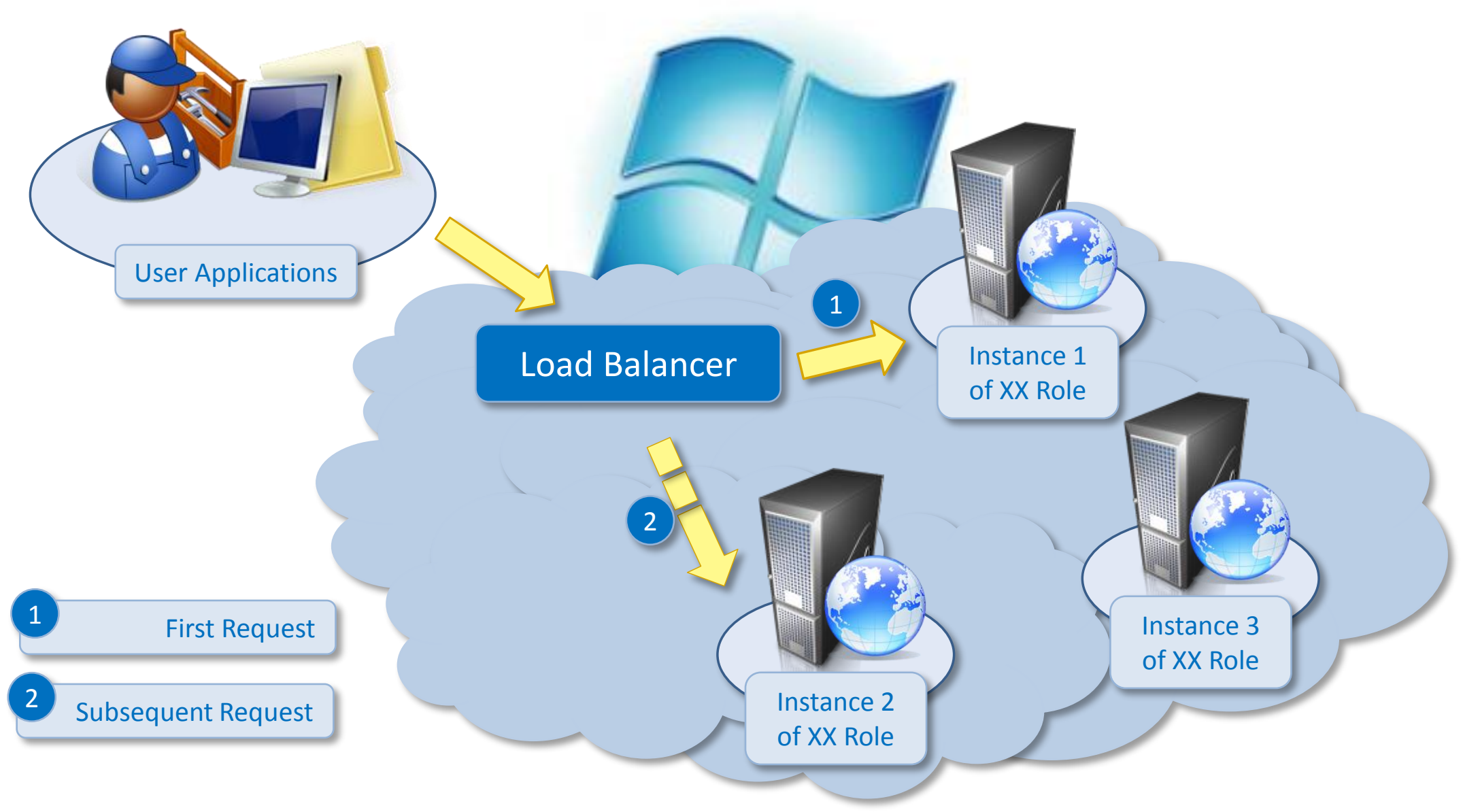

***Figure 6: Routing in Windows Azure.***

As depicted in **Figure 6**, instances from the same Role will share the same defined input endpoints and be behind the same Load Balancer. It is the responsibility of the Load Balancer to dispatch the incoming external traffic to instances behind it following a round robin mechanism. For instance, the first request will be sent to the first instance of the given worker role, the second will be sent to the second available instance, and so forth.

As we plan to deploy Aneka Container as instances of the Windows Azure Worker Role, there exists a situation where the Aneka Master is outside the Windows Azure Cloud and tries to send messages to a specific Aneka Worker inside the Windows Azure Cloud. Since the load balancer is responsible for forwarding these messages, there is a good possibility that the message may be sent to a Aneka Worker Container other than the specified one. Hence, in order to avoid the message being transferred to the wrong Aneka Worker Container, two possible solutions are available:

- **Forward Messages among Aneka Worker Containers.** When a Container receives a message that does not belong to it, it will forward the message to the right Container according to the *InternalEndpoint* address encoded in the *NodeURI* of Target Node of the Message. The advantage of this solution is the consistency of the architecture of Aneka PaaS since no new components are introduced to the architecture. The disadvantage, however, is that the performance of Aneka Worker Containers will be hindered due to the overhead of forwarding message.
- **Deploy a message proxy between the Aneka Worker Containers and Master for the purpose of dispatching incoming external messages.** The Message Proxy is a special kind of Aneka Worker Container which does not host any execution services. When the Windows Azure Aneka Cloud starts up, all Aneka Worker Containers in Windows Azure encode the internal endpoint address into the *NodeURI*. When the Message Proxy receives a message from the Master, it dispatches the message to the right Aneka Worker Container according to the encoded *NodeURI* specified in the Target Node of Message. The disadvantage of this solution is that it costs extra since Windows Azure charges according to the number of instances launched. However, in view of possible performance issues, the second solution is preferred. More details on the deployment of the Message Proxy Role are introduced in **Section 3.1.3.**

### 2.6.3 Dynamic Environment

As mentioned in **Section 2.6.2**, each Windows Azure Web Worker Role can define an external endpoint to listen on a unique port. As a matter of fact, the port of endpoints defined is the port of the load balancer. The Windows Azure Fabric will dynamically assign a random port number to each instance of the given role to listen on. Consequently, before starting the Container, we need to get the dynamically assigned endpoint via *RoleEnvironment.CurrentRoleInstance.InstanceEndpoints* Property defined in the *Windows Azure Managed Library* and save it to the Aneka Configuration File so that the Container can bind the TCP channel to the right port.

Another change required by the dynamic environment of Windows Azure is that we need to set the *NodeURI* of an Aneka Worker Container to the URL of the Message Proxy and encode the internal endpoint of the Container into the URL. When the Aneka Master sends a message to the *NodeURI* of an Aneka Worker Container, the Message Proxy receives the message and forwards it to the right Aneka Worker Container according to the internal endpoint address encoded in the *NodeURI*.

Furthermore, due to the dynamic nature of Windows Azure, we also need to guarantee that the Load Balancer sends the message to the instance of Message Proxy Role only if the message channel of the instance is ready and all the instances of Aneka Worker Role start to send Heartbeat Message to the Aneka Master located on-premises, after the deployment of the Message Proxy Role is finished.

### 2.6.4 Debugging

Debugging a Windows Azure Application is a bit different from debugging other Windows .Net applications.

In general, after we install *Windows Azure Tools for Visual Studio* we can debug a Windows Azure application locally when it is running in the Development Fabric during the development stage. However, after the application has been deployed on Windows Azure public Cloud, we cannot remotely debug the deployed application since we do not have direct access and administrative privilege on Windows Azure VMs.Fortunately, in June 2010 Windows Azure Tools + SDK, Windows Azure provides us with a new feature that enables us to debug issues that occur in the Cloud via *IntelliTrace*. With *IntelliTrace* debugging we can log extensive debugging information for a role instance while it is running in Windows Azure. Subsequently, we can use the *IntelliTrace* logs to step through the code from Visual Studio.

# 3. DESIGN

In this section, we will discuss the design decisions for deploying Aneka Containers on Windows Azure as instances of Worker Role, how to integrate and leverage the dynamic provisioning service of Aneka, and how to exploit the Windows Azure Storage as a file storage system for Aneka PaaS in detail. The deployment includes two different types. The first type is to deploy Aneka Worker Containers on Windows Azure while the Aneka Master Container is run on local or on-premise resource. The second type is to deploy the entire Aneka PaaS including Aneka Master Container and Aneka Worker Containers on Windows Azure.

## 3.1 Deploying Aneka Workers on Windows Azure

### 3.1.1 Overview

**Figure 7** provides an overall view of the deployment of Aneka Worker Containers as instances of Windows Azure Worker Role.

As shown in the figure, there are two types of Windows Azure Worker Roles used. These are the **Aneka Worker Role** and **Message Proxy Role**. In this case, we deploy one instance of Message Proxy Role and at least one instance of Aneka Worker Role. The maximum number of instances of the Aneka Worker Role that can be launched is limited by the subscription offer of Windows Azure Service that a user selects. In the first stage of the project, the **Aneka Master Container** will be deployed in the on-premises private Cloud, while **Aneka Worker Containers** will be run as instances of Windows Azure Worker Role. The instance of the Message Proxy Role is used to transfer the messages sent from the Aneka Master to the given Aneka Worker.

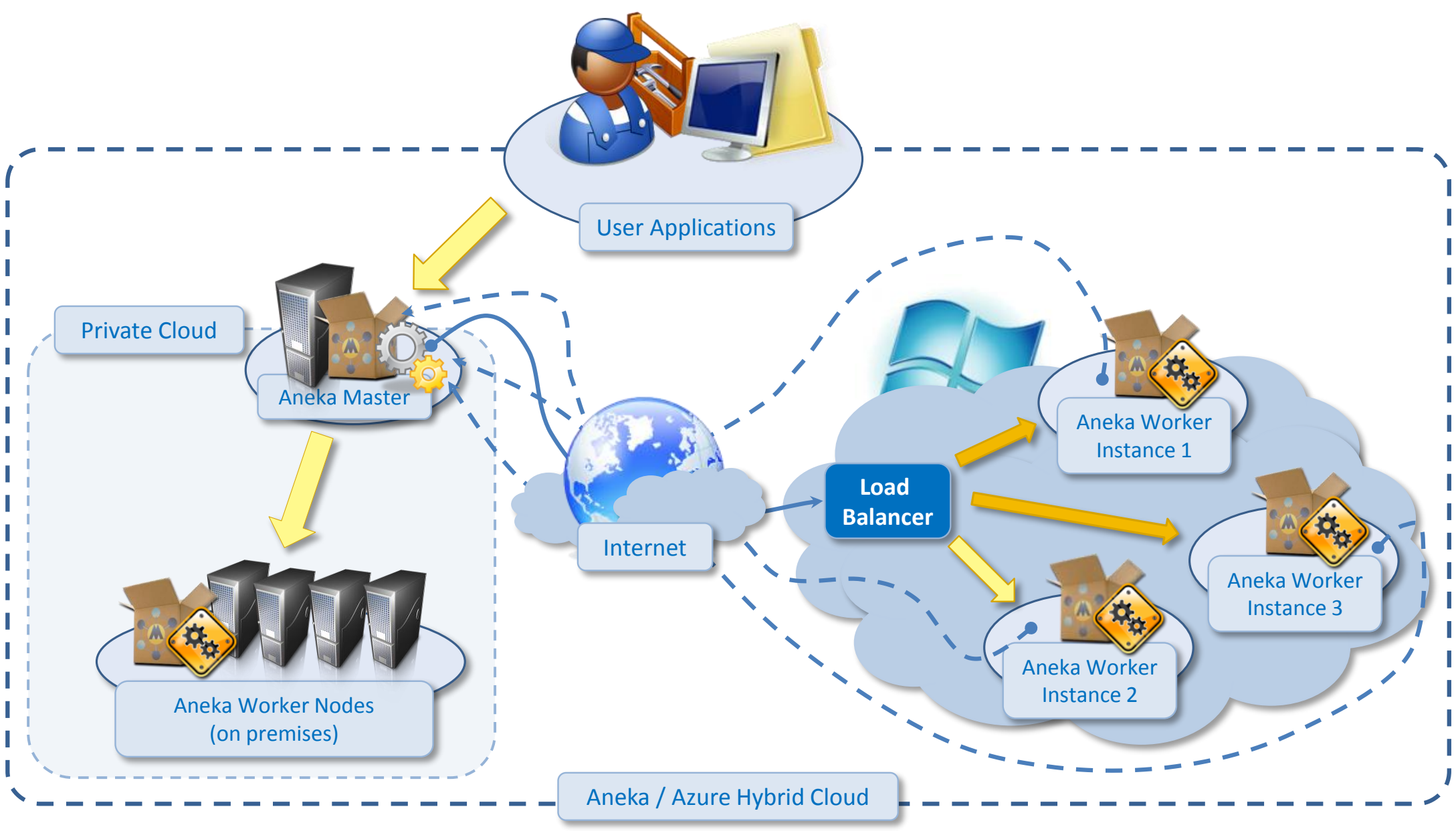

***Figure 7: The Deployment of Aneka Worker Containers as Windows Azure Worker Role Instances.***

In this deployment scenario, when a user submits an application to the Aneka Master, the job units will be scheduled by the Aneka Master by leveraging on-premises Aneka Workers, if they exist, and Aneka Worker instances on Windows Azure simultaneously. When Aneka Workers finish the execution of Aneka work units, they will send the results back to Aneka Master, and then Aneka Master will send the result back to the user application.

### 3.1.2 Aneka Worker Deployment

Basically, we can deploy Aneka Containers of the same configuration as an Azure Worker Role, since they share the same binary code and the same configuration file. We can setup the number of instances of an Azure Worker Role to be launched in the Windows Azure Service Configuration file, which represents the number of Aneka Containers that will be deployed on the Windows Azure Cloud. And also we need to setup the Aneka Master URI and the shared security key in the Windows Azure Service Configuration file. When the instances of Aneka Worker Role are started up by Windows Azure Role Hosting Process, we firstly update the configuration of the Aneka Worker Container and start the Container program. After the container starts successfully, it will connect to the Aneka Master directly.

### 3.1.3 Message Proxy

As for the issue we discussed in **Section 2.6.2**, in order to guarantee that messages are transferred to the right target node specified by the Aneka Master, we need a mechanism to route messages to a given instance. Therefore, we introduce a Message Proxy between the Load Balancer and Aneka Worker instances. As shown in the **Figure 8**, all the messages that are sent to the Aneka Worker Containers in Windows Azure Public Cloud will be transferred to the external input endpoint of Message Proxy Role. All the messages will be transferred to the load balancer of the input endpoint①. The load balancer will transfer the messages to the instance of Message Proxy Role②. In this case, we only launch one instance for Message Proxy Role. The Message Proxy picks the incoming message, and parses the *NodeURI* of target node to determine the internal address of the target node, and then forwards the messages to the given Aneka Worker③. The Aneka Worker will handle the message and send a reply message to Aneka Master directly④.

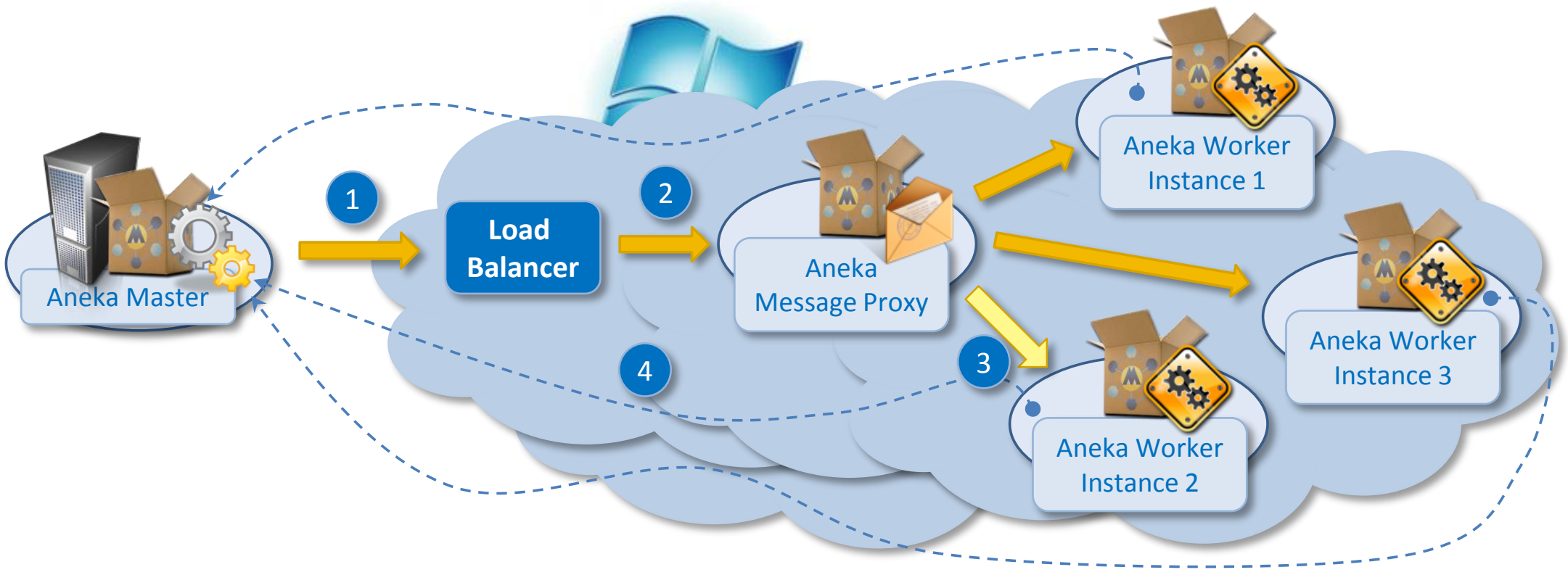

***Figure 8: How the Message Proxy Works.***

### 3.1.4 Dynamic Provisioning

Windows Azure provides us with programmatic access to most of the functionality available through the Windows Azure Developer Portal via the *Windows Azure Service Management REST API*. Using the Service Management API, we can manage our storage accounts and hosted services.

Hence, by using the completely extensible *Dynamic Resource Provisioning Object Model* of Aneka PaaS and *Windows Azure Service Management REST API*, we can integrate Windows Azure Cloud resources into Aneka's infrastructure and provide support for dynamically provisioning and releasing Windows Azure resource on demand.

Specifically, the Aneka APIs offer the *IResourcePool* interface and the *ResourcePoolBase* class as extension points for integrating new resource pools. By implementing the interface and extending the abstract base class, we can support provisioning of Aneka Worker Containers on Windows Azure by following these steps:

- Use the *CSPack Command-Line Tool* to programmatically packet all the binaries and the service definition file to be published to the Windows Azure fabric, into a service package file;
- Use the *Windows Azure Storage Services REST API* to upload the service package to Windows Azure Blob;
- Use the *Windows Azure Service Management REST API* to create, monitor and update the status of the deployment of the Windows Azure Hosted Service.
- Use the *Windows Azure Service Management REST API* to increase or decrease the number of instances of Aneka Worker Containers to scale out or scale in on demand;
- Use the *Windows Azure Service Management REST API* to delete the whole deployment of the Windows Azure Hosted Service when the provisioning service is shutting down.

## 3.2 Deploying Aneka Cloud on Windows Azure

In the second deployment scenario, we deploy the Aneka Master Container as instance of Windows Azure Worker Role. After finishing this step, we can run the whole Aneka Cloud infrastructure on the Windows Azure Cloud Platform, as can be seen from **Figure 9**.

### 3.2.1 Overview

In this scenario, users submit Aneka applications outside of the Windows Azure Cloud and receive the result of the execution from Windows Azure Cloud. The advantage of this structure is that it can dramatically decrease message transfer delay since all the messages between the Aneka Master and Aneka Workers are transferred within the same data centre of Windows Azure, and the cost of data transfer charged by Windows Azure will reduce greatly as well.

Further, for data persistence requirements, the Aneka Master Container, can directly use the Relational Data Service Provided by SQL Azure which would have higher data transfer rates and of higher security since they are located in the same Microsoft data centre.

### 3.2.2 Aneka Deployment in Azure

**Figure 9** shows two types of roles being deployed on the Windows Azure Cloud: one instance of Aneka Master Role hosting the Aneka Master Container, and at least one instance of the Aneka Worker Role hosting the Aneka Worker Container. The Aneka Master Container and Aneka Worker Containers interact with each other via an internal endpoint, whilst the client and Aneka Master Container interact via an external endpoint of the Aneka Master instance.

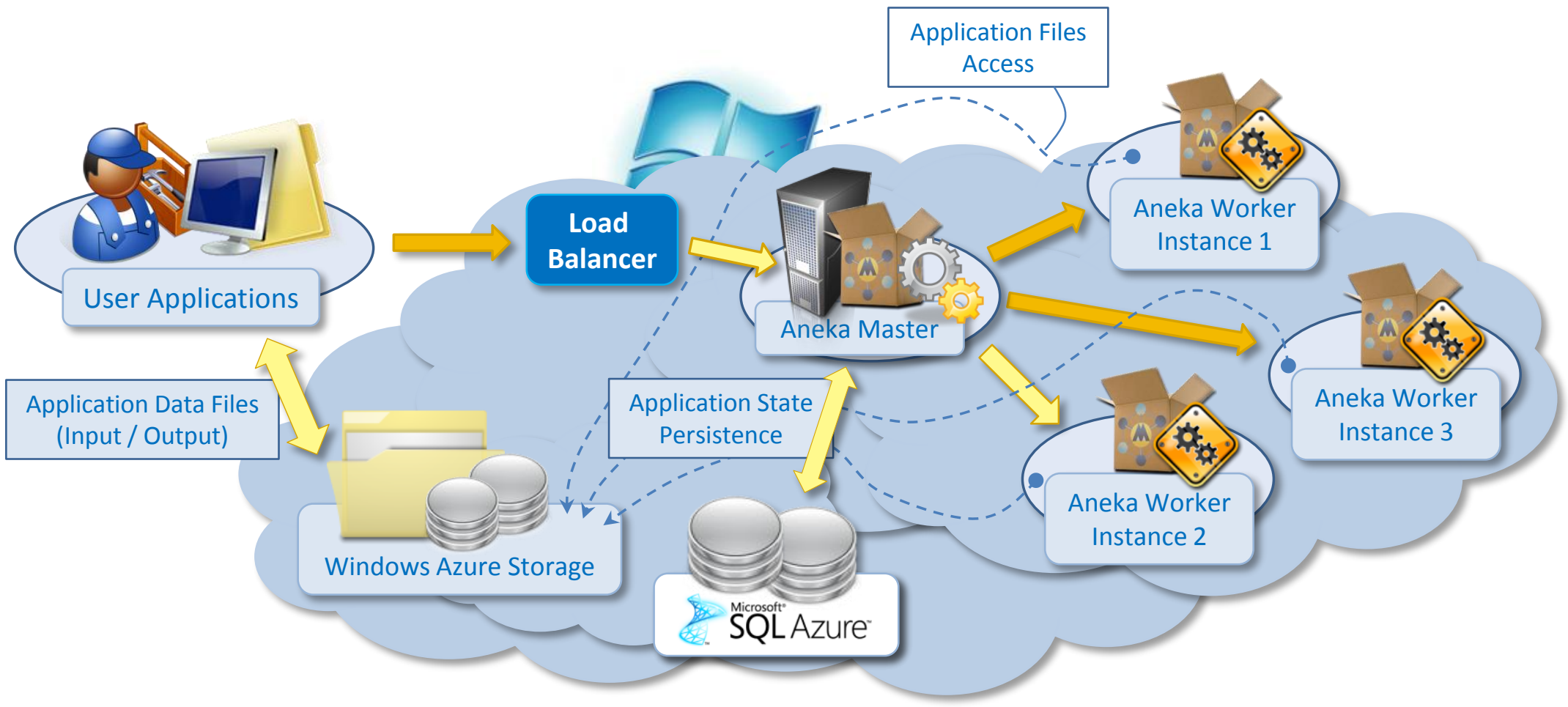

*Figure 9: The Deployment of Aneka Master Container.*

### 3.2.3 File Transfer System

In the current version of Aneka Cloud, the FTP protocol is used to transfer data files from the client to the Aneka Master (or a separate Storage Container) and between the Aneka Master and Aneka Worker Containers. However, due to the limitation of a maximum of 5 networking ports allowed on each Windows Azure Role instance, we can no longer use the FTP service to support file transfers on the Windows Azure Cloud. Instead, we can leverage Windows Azure Storage to support file transfers in Aneka.

In general, as illustrated in **Figure 10**, two types of Windows Azure Storage will be used to implement the Aneka File Transfer System: Blobs and Queues. Blobs will be used for transferring data files, and Queues for the purpose of notification. When Aneka users submit the application, if the transfer of input data files is needed, the *FileTransferManager* component will upload the input data files to the Windows Azure Blob and notify the start and end of the file transfer to Aneka's Storage Service via Windows Azure Queue. Similarly, the Aneka Worker will download the related input data file from Windows Azure Blob, and the start and end of the file transfer will be notified via Windows Azure Queue. When the execution of the work unit is completed in the Aneka Worker, if the transfer of output data files is needed, the *FileTransferManager* component of Aneka PaaS will upload the output data files to the Windows Azure Blob to enable Aneka users to download from it.

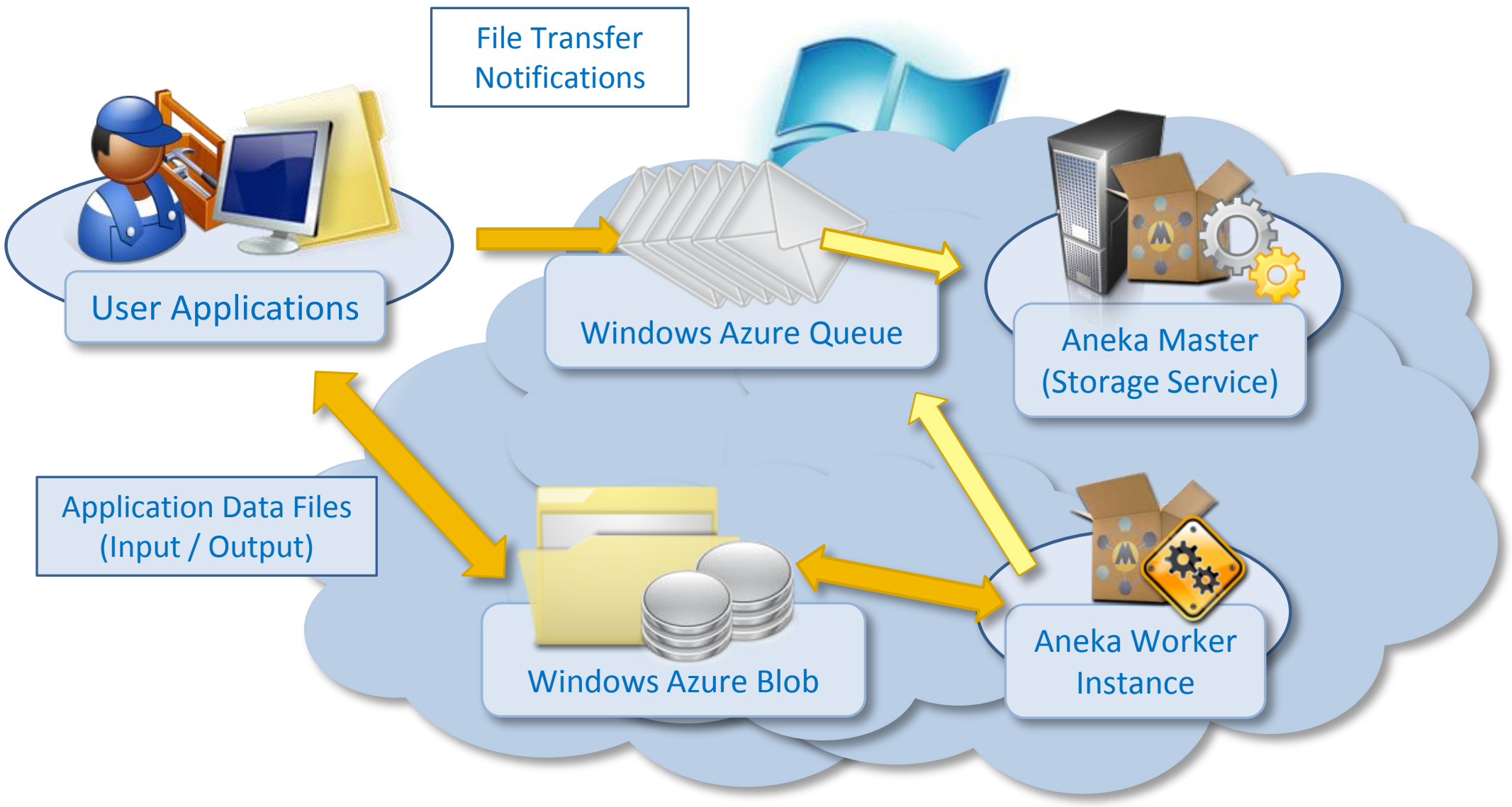

***Figure 10: Using Windows Azure Storage Blob and Queue for Implementation of Aneka File Transfer System.***

## 4. IMPLEMENTATION

In this section, we will explore the implementation details of the design we presented in Section 3. Section 4.1 displays the class diagrams of the new and changed components in Aneka PaaS. Next, Section 4.2 illustrates the configuration setting of the deployments, whilst Section 4.3 demonstrates the designed life cycle of deployments.

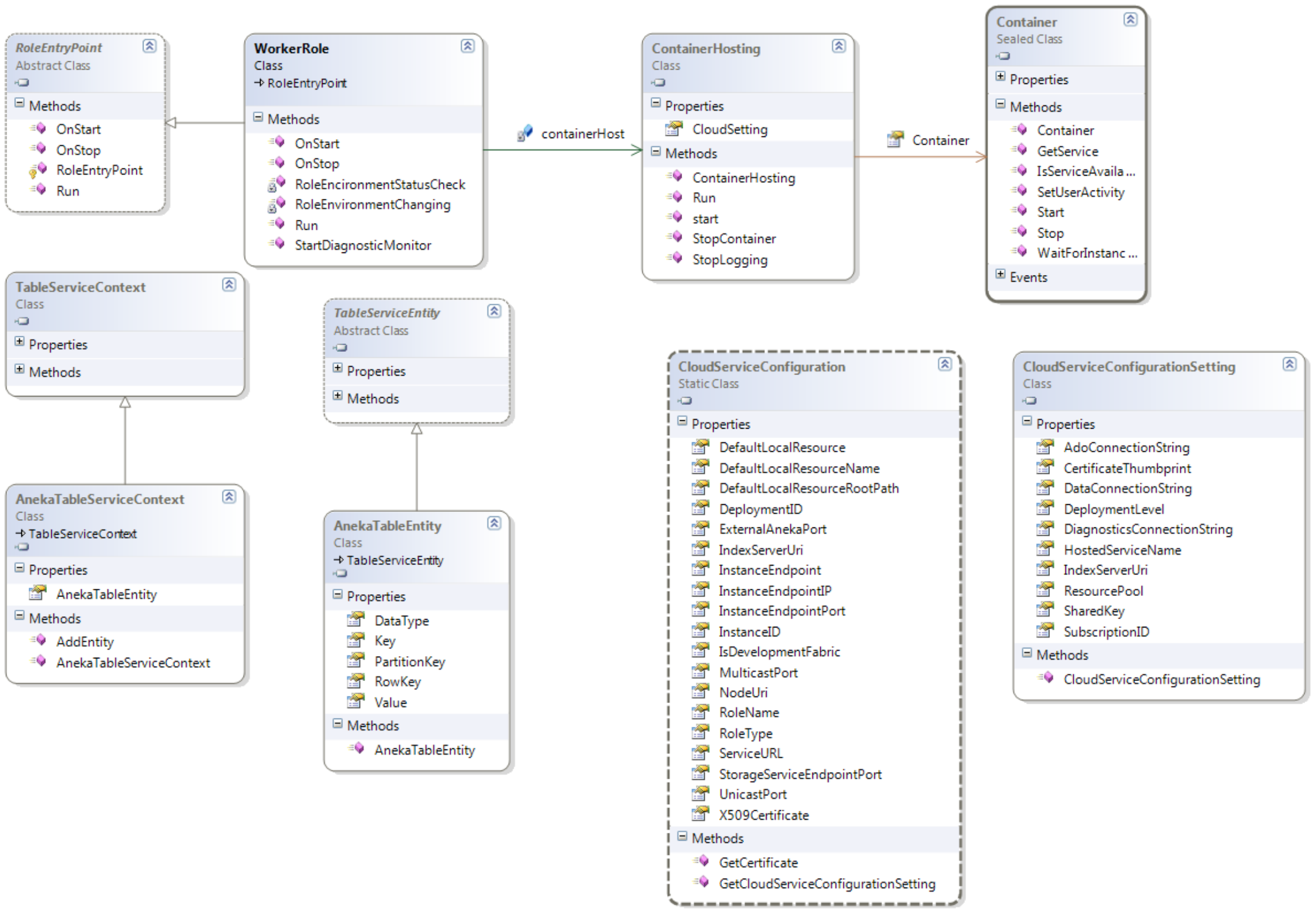

***Figure 11: Class Diagram for Windows Azure Aneka Container Hosting Component.***

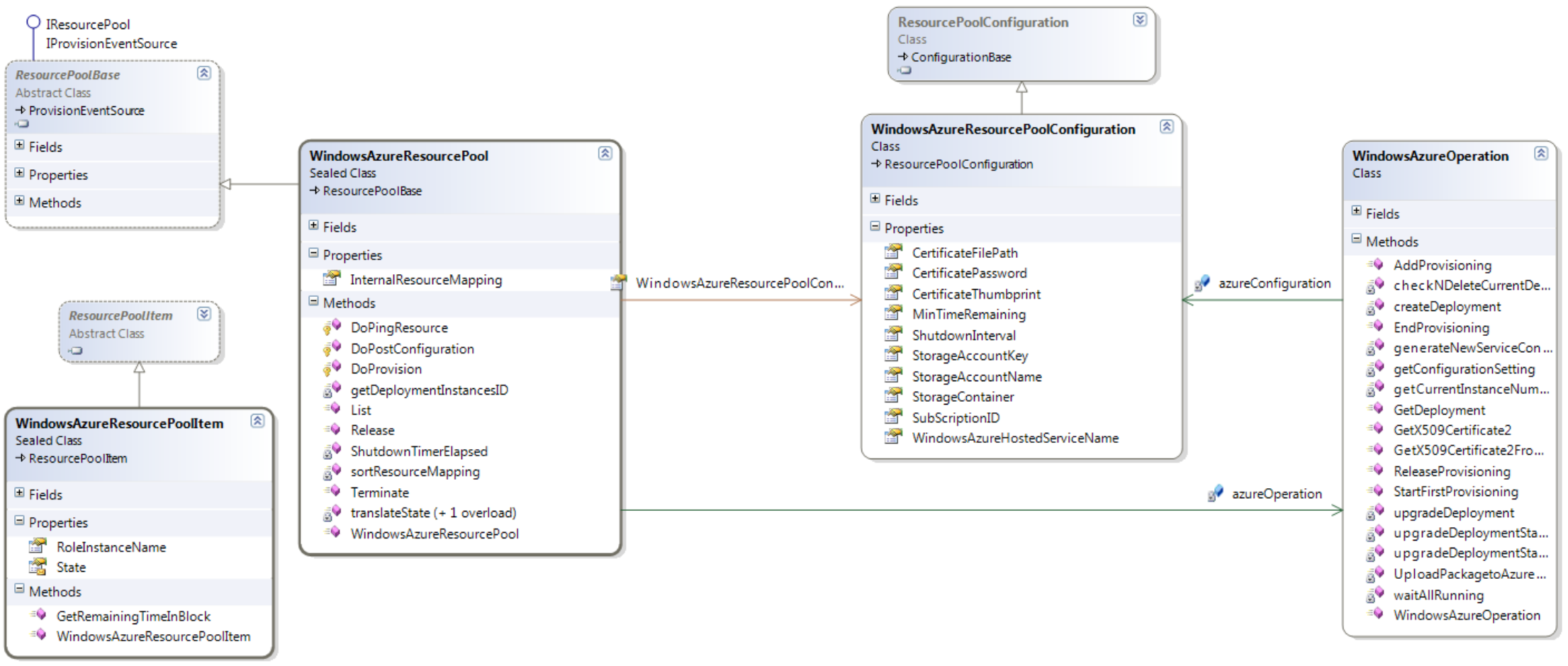

***Figure 12: Class Diagram for Windows Azure Aneka Provisioning Resource Pool Component.***

## 4.1 Class Diagrams

### 4.1.2 Windows Azure Aneka Container Deployment

Technically, in order to start an Aneka Container on Windows Azure Role instance, we need to extend the *RoleEntryPoint* class which provides a callback to initialize, run, and stop instances of the role. We override the Run() method to implement our code to start the Aneka Container which will be called by Windows Azure Runtime after the role instance has been initialized. Also worth noting is that due to the dynamic nature of the Windows Azure environment, the configuration of Aneka

Worker Containers must be updated using the information obtained from the *CloudServiceConfiguration* Class.

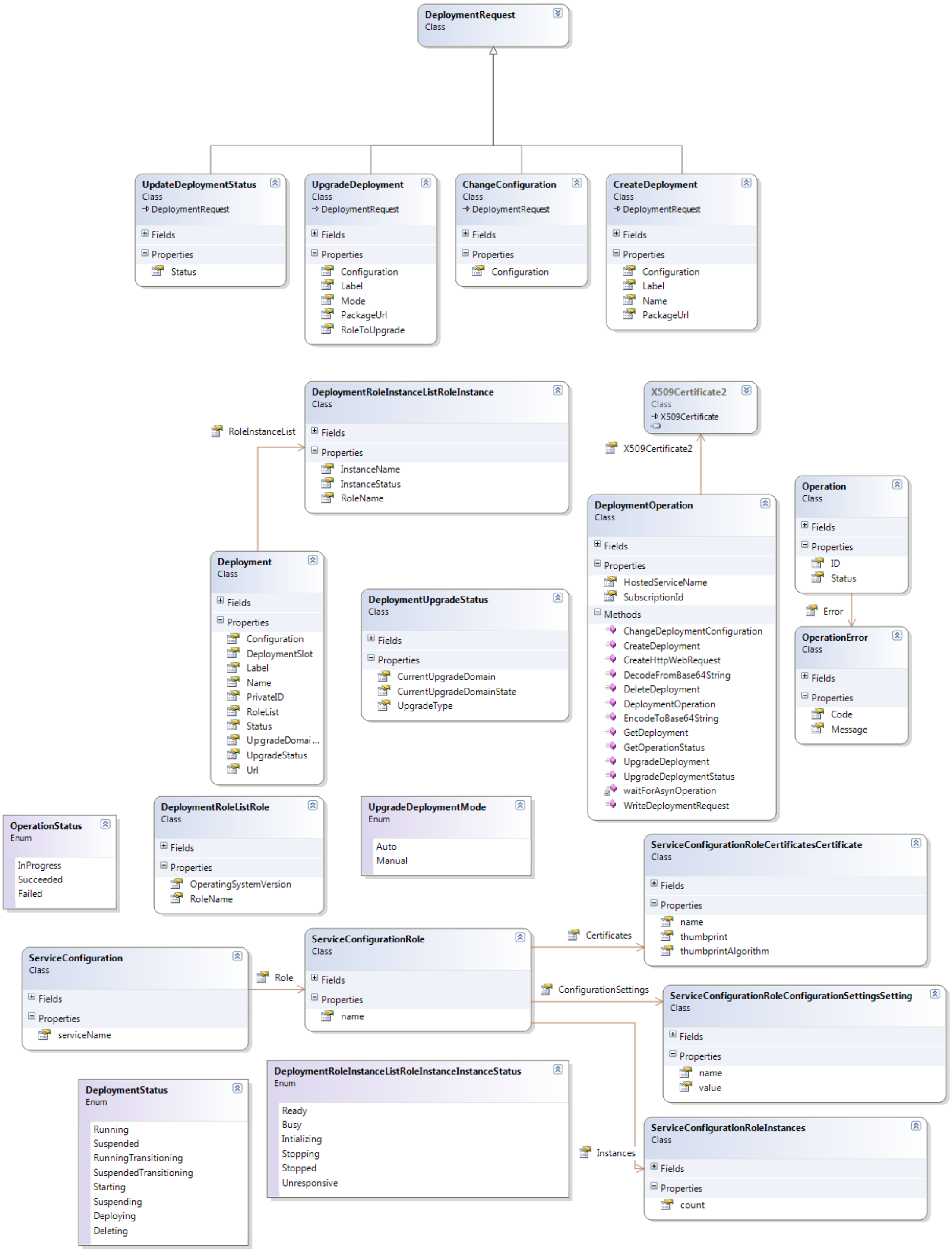

***Figure 13: Class Diagram for Windows Azure Service Management Component.***

## 4.1.2 Windows Azure Provisioning Resource Pool

The extendable and customizable Dynamic Resource Provisioning Object Model of Aneka PaaS enables us to provide new solutions for dynamic provisioning in Aneka.

Specifically speaking, the *WindowsAzureResourcePool* class extends the *ResourcePoolBase* class and implements the *IResourcePool* interface to integrate Windows Azure as a new resource pool. The class *WindowsAzureOperation* provides all the operations that are needed to interact with the Windows Azure Service Management REST API.

### 4.1.3 Windows Azure Service Management

The *DeploymentOperation* component is used to interact with the Windows Azure Service Management REST API to manage the Windows Azure Hosted Services in terms of creating a deployment, updating the status of a deployment (such as from *Suspended* to *Running* or vice versa) upgrading the deployment, querying the deployment and deleting the deployment. This component is used by the Resource Provisioning Service to manage the Windows Azure resource pool, and is also used by the Windows Azure Role Deployment to monitor the status of deployment.

### 4.1.4 File Transfer System

The File Transfer System Component is used to transfer data files which are used in application between clients and Aneka Cloud deployed on top of Windows Azure. The class *AzureFileChannelController* which implements the *IFileChannelController* interface represents the server component of the communication channel. It is responsible for providing the connection string for the client component to gain access to the Windows Azure Storage Service providing a way to upload and retrieve a specific file. The class *AzureFileHandler* which implements the *IFileHandler* interface is in charge of retrieving a single file or a collection of files from the server component of the communication channel and uploading a single file or a collection of files to the server component of the communication channel.

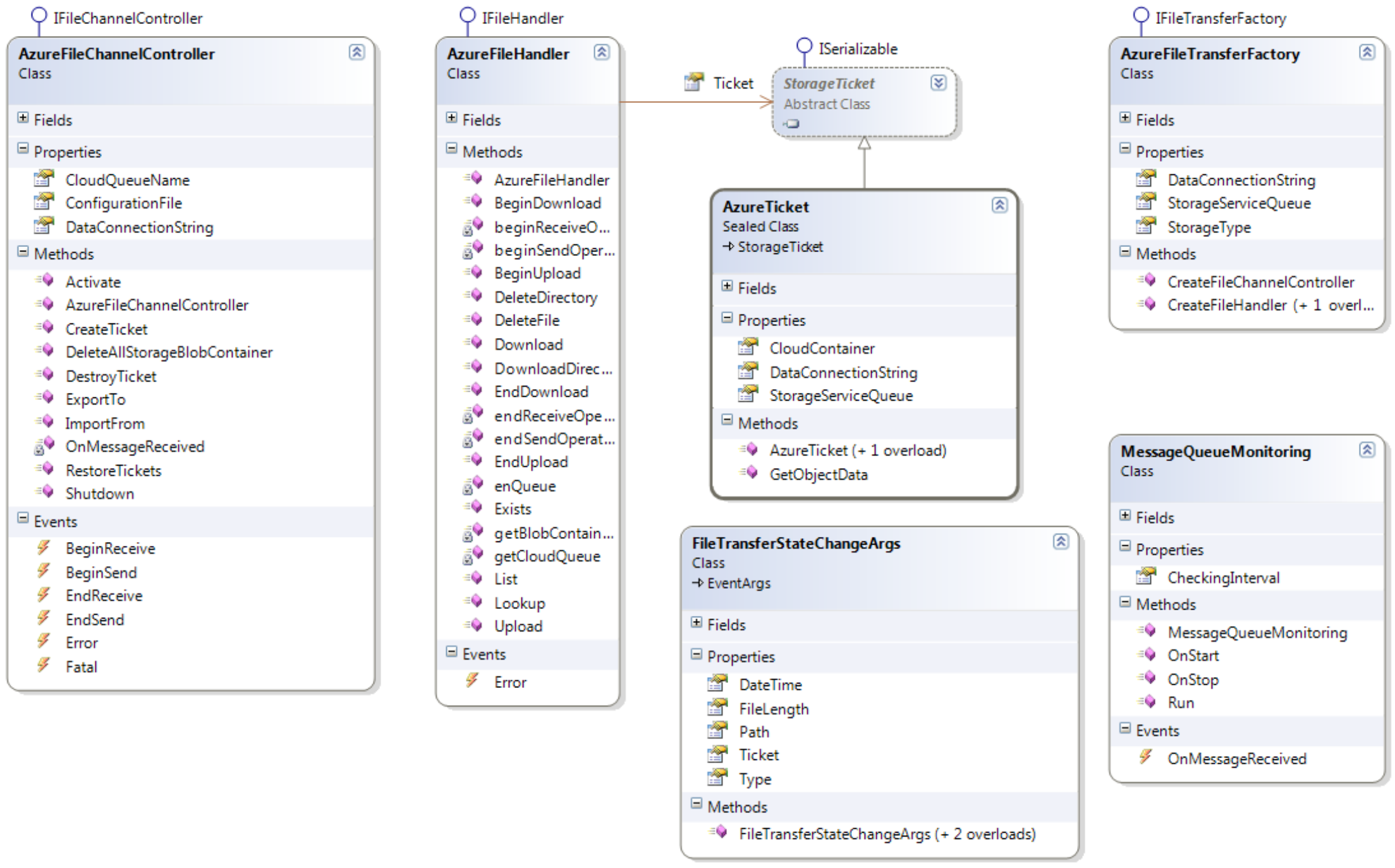

***Figure 14: Class Diagram for Windows Azure Aneka Storage Service Implementation using Windows Azure Storage.***

## 4.2 Configuration

### 4.2.1 Provisioning Aneka Workers from Aneka Management Studio

In order to enable the Provisioning Service of Aneka to provision resources on Windows Azure, we need to configure it via the Aneka Cloud Management Studio, while configuring the services of the Aneka Master Container. This requires configuring the Scheduling Service and Resource Provisioning Service.

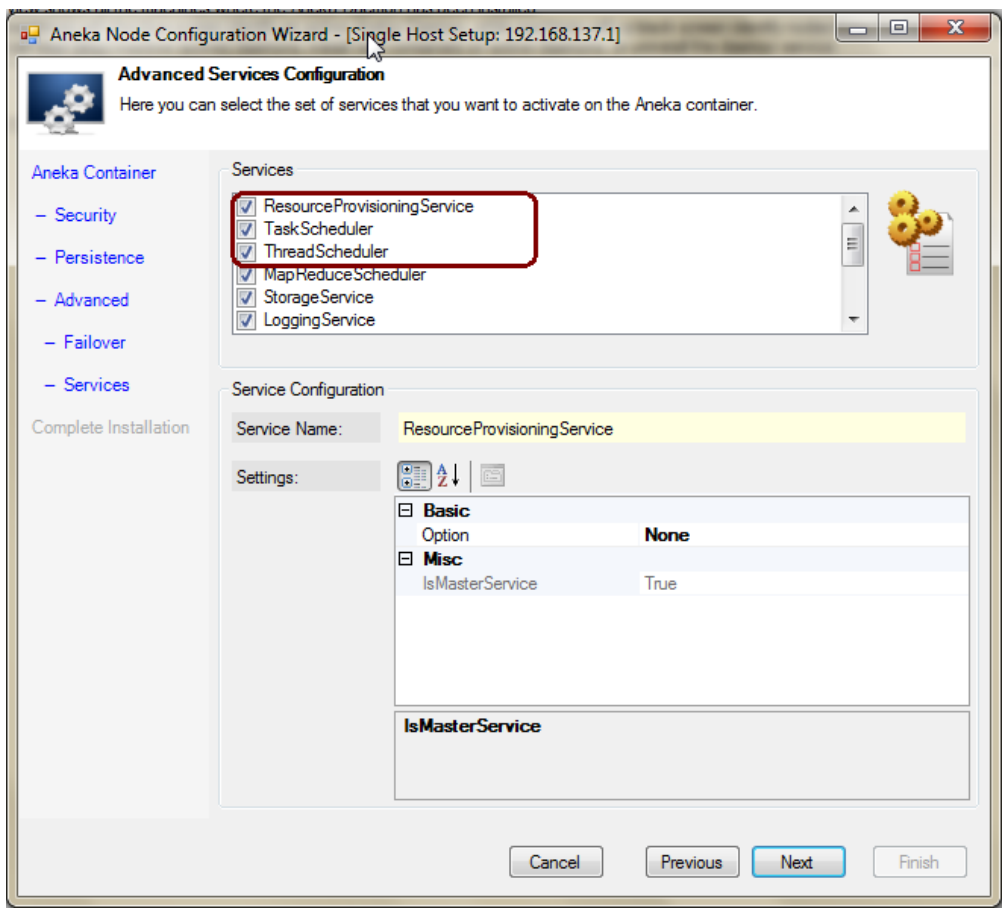

*Figure 15: Provisioning Service Configuration.*

For the Scheduling Service, we need to select a proper scheduling algorithm for the *TaskScheduler* and *ThreadScheduler*. Currently, only two algorithms are available for dynamic provisioning: *FixedQueueProvisioningAlgorithm* and *DeadlinePriority-ProvisioningAlgorithm*.

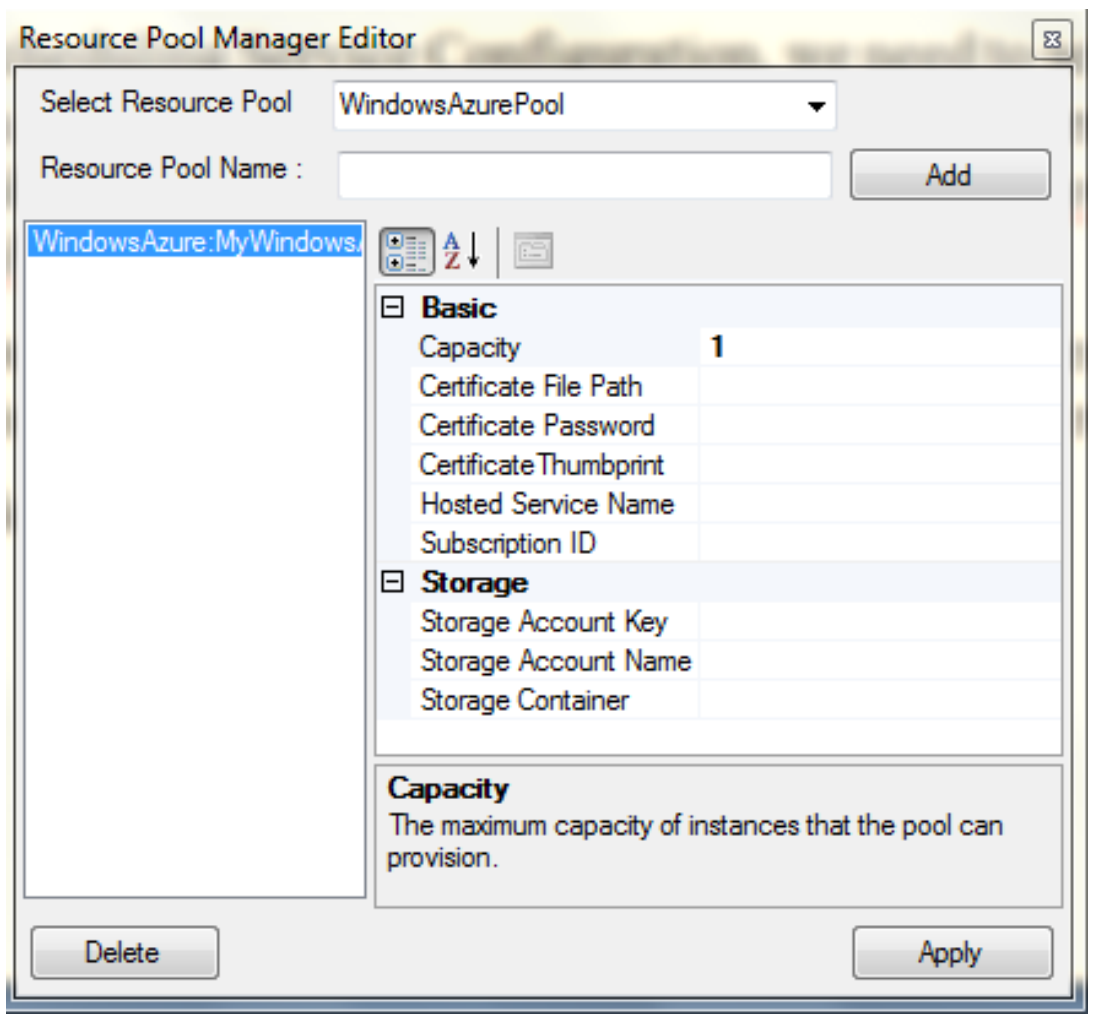

*Figure 16: Windows Azure Resource Provisioning Service Configuration.*

The configuration required for the Resource Provisioning Service in order to acquire resources from the Windows Azure Cloud Service Providers is depicted in **Figure 16**. For setting up a Windows Azure Resource Pool, we need the following information:

- **Capacity:** identifies the number of instances that can be managed at a given time by the pool. This value is restricted to the maximum number of instances that the user is allowed to launch on Windows Azure, based on the subscription.

- **Certificate File Path:** specifies the file path of an X509 certificate that is used to interact with Windows Azure Service Management REST API. This certificate must also be uploaded to the *Windows Azure Development Portal*.
- **Certificate Password:** designates the password of the X509 Certificate.
- **Certificate Thumbprint:** assigns the thumbprint of the X509 Certificate.
- **Hosted Service Name:** identifies the name of the Windows Azure Hosted Service; the service must have been created via the *Windows Azure Development Portal*.
- **Subscription ID:** specifies the Subscription ID of Windows Azure Account.
- **Storage Account Name:** designates the name of Windows Azure Storage account that is under the same subscription.
- **Storage Account Key:** specifies the key of the storage account.
- **Storage Container:** defines the name of storage container which is used to store the Windows Azure Hosted Service Package File.

```xml
<?xml version="1.0"?>
<ServiceConfiguration serviceName="AnekaOnWindowsAzure" xmlns="http://schemas.microsoft.com/ServiceHosting/2008/10/ServiceConfigu
  <Role name="AnekaMaster">
    <Instances count="1" />
    <ConfigurationSettings>
      <Setting name="DiagnosticsConnectionString" value="DefaultEndpointsProtocol=https;AccountName=anekacloud;AccountKey=eGhmuL1
      <Setting name="DataConnectionString" value="DefaultEndpointsProtocol=https;AccountName=anekacloud;AccountKey=eGhmuL194C9QA+
      <Setting name="SharedKey" value="Qq6dthHKWph0QkS5X7rJL0qLeR14IQfgMexGapTBouijEZzy2XGM3ytK/uldFHQB" />
      <Setting name="IndexServerUri" value="tcp://localhost:3333/Aneka" />
      <Setting name="ResourcePool" value="MyWindowsAzurePool" />
      <Setting name="SubscriptionID" value="a22fc8fe-5955-421f-a370-75e3f1246323" />
      <Setting name="HostedServiceName" value="anekacloud" />
      <Setting name="CertificateThumbprint" value="81841B188C32BE42B5256CAED1CE905099785CA9" />
      <Setting name="DeploymentLevel" value="Master" />
      <Setting name="AdoConnectionString" value="Server=tcp:w1fypjjbpf.database.windows.net;Database=aneka;User ID=aneka@w1fypjjt
    </ConfigurationSettings>
    <Certificates>
      <Certificate name="SelfManagement" thumbprint="81841B188C32BE42B5256CAED1CE905099785CA9" thumbprintAlgorithm="sha1" />
    </Certificates>
  </Role>
  <Role name="AnekaWorker">
    <Instances count="5" />
    <ConfigurationSettings>
      <Setting name="DiagnosticsConnectionString" value="DefaultEndpointsProtocol=https;AccountName=anekacloud;AccountKey=eGhmuL1
      <Setting name="DataConnectionString" value="DefaultEndpointsProtocol=https;AccountName=anekacloud;AccountKey=eGhmuL194C9QA+
      <Setting name="SharedKey" value="Qq6dthHKWph0QkS5X7rJL0qLeR14IQfgMexGapTBouijEZzy2XGM3ytK/uldFHQB" />
      <Setting name="IndexServerUri" value="tcp://localhost:3333/Aneka" />
      <Setting name="ResourcePool" value="MyWindowsAzurePool" />
      <Setting name="SubscriptionID" value="a22fc8fe-5955-421f-a370-75e3f1246323" />
      <Setting name="HostedServiceName" value="anekacloud" />
      <Setting name="CertificateThumbprint" value="81841B188C32BE42B5256CAED1CE905099785CA9" />
      <Setting name="DeploymentLevel" value="Master" />
      <Setting name="AdoConnectionString" value="" />
    </ConfigurationSettings>
    <Certificates>
      <Certificate name="SelfManagement" thumbprint="81841B188C32BE42B5256CAED1CE905099785CA9" thumbprintAlgorithm="sha1" />
    </Certificates>
  </Role>
```

***Figure 17: Windows Azure Service Configuration File related to Windows Azure Aneka Cloud Package.***

### 4.2.2 Deploying Aneka Cloud on Windows Azure

In order to deploy an Aneka Cloud on Windows Azure, before uploading the Windows Azure Aneka Cloud Package into Windows Azure Cloud, we need to configure the Windows Azure Service Configuration file related to the Windows Azure Aneka Cloud Package. To be more specific, as shown in the **Figure 17**, we need to specify the values below:

- **DiagnosticsConnectionString:** The connection string for connecting to the Windows Azure Storage Service which is used to store diagnostics data.

- **DataConnectionString:** The connection string for connecting to Windows Azure Storage Service which is used to implement the File Transfer System.
- **SharedKey:** The security key shared between Aneka Master and Aneka Worker.
- **SubscriptionID:** The Subscription ID of Windows Azure Account.
- **HostedServiceName:** The name of the Windows Azure Hosted Service.
- **CertificateThumbprint:** The thumbprint of the X509 Certificate which has been uploaded to Windows Azure Service Portal. The value of thumbprint in the Certificate Property should also be set.
- **AdoConnectionString:** The connection string used to connect to an ADO relational database if relational database is used to store persistent data.

More importantly, we need to define the **Instance Number** of Aneka Workers running on Windows Azure Cloud, which is specified in the "count" attribute of "Instance" property.

## 4.3 Life Cycle of Deployment

### 4.3.1 Aneka Worker Deployment

**Figure 18** shows the whole life cycle of deployment of Aneka Worker Containers on Windows Azure Cloud.

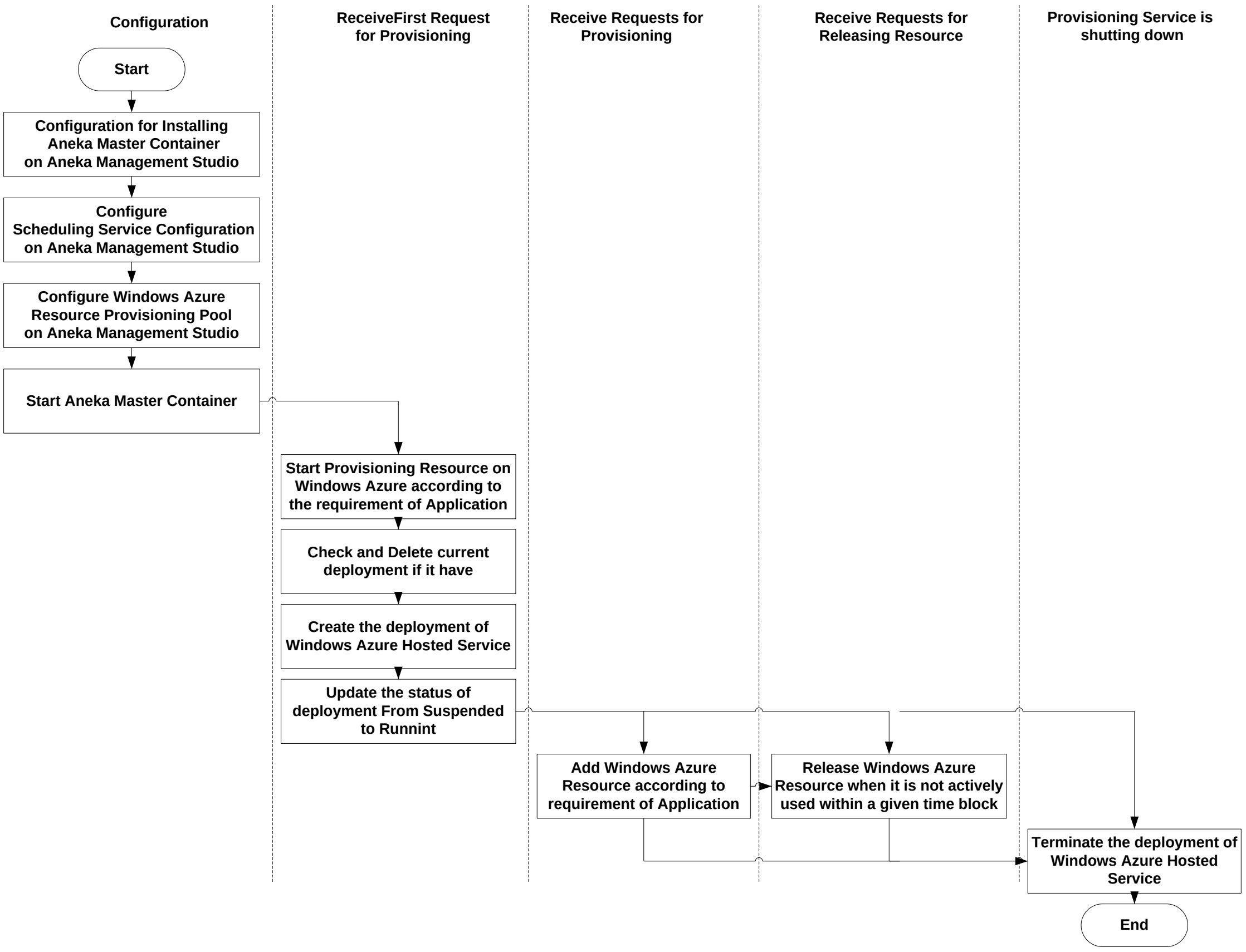

*Figure 18: The Life Cycle of Aneka Worker Container Deployment on Windows Azure.*

Generally speaking, the whole life cycle of Aneka Worker Container deployment on Windows Azure involves five steps. They are *Configuration*, *First Time Resource*

*Provisioning*, *Subsequent Resource Provisioning*, *Resource Release*, and *Termination of Deployment* respectively.

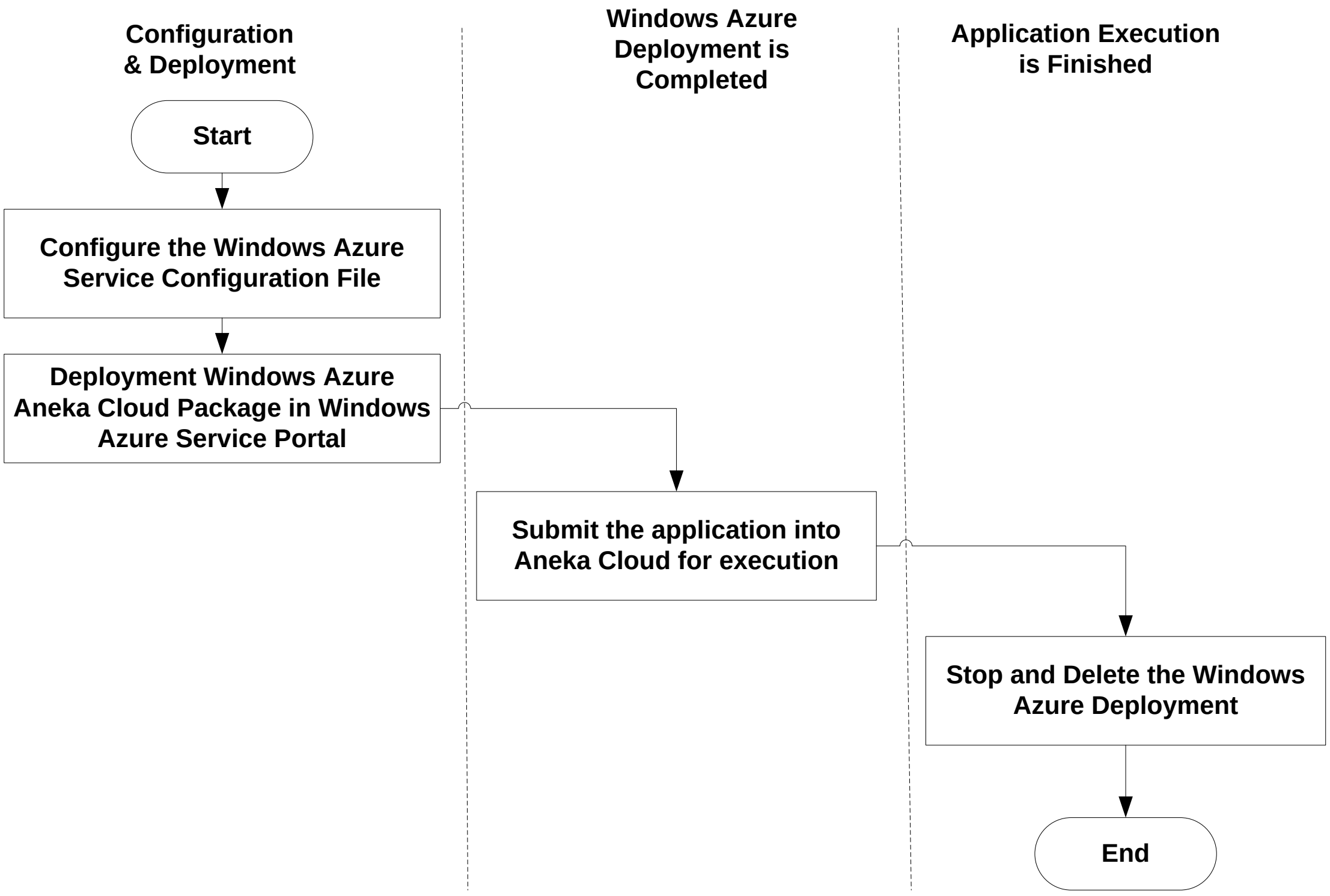

***Figure 19: The Life Cycle of Aneka Cloud Deployment on Windows Azure.***

### 4.3.2 Aneka Cloud Deployment

**Figure 19** shows the life cycle for deploying an entire Aneka Cloud on top of Windows Azure.

In general, the whole life cycle of Aneka Cloud deployment on Windows Azure involves 3 steps. They are *Configuration* and *Deployment*, *Application Execution*, and *Deployment Termination* respectively.

## 5. EXPERIMENTS

In this section, we will present the experimental results for application execution on the Aneka Windows Azure Deployments including Aneka Worker Deployment and Aneka Cloud Deployment. The test application we selected is Mandelbrot application (**Figure 20**) which is developed on top of Aneka Thread Model to determine the suitability of Aneka Windows Azure Deployment for running parallel algorithms.

**Figure 21** displays the experimental results for executing the Mandelbrot application using different input problem sizes, running on both Aneka Worker Deployment and Aneka Cloud Deployment on Windows Azure when the number of Aneka Workers being launched is 1, 5, and 10. The compute instance size of the Azure Instance selected to run the Aneka Worker Containers is *small computer instance* which is a virtual server with dedicated resources (CPU 1.6 GHz and Memory 1.75 GB) and specially tailored for Windows Server 2008 Enterprise operating system as the guest OS. The instance size for deploying the Aneka Master Container is *medium computer instance* with machine configuration CPU 2*1.6 GHz and Memory 3.5 GB.

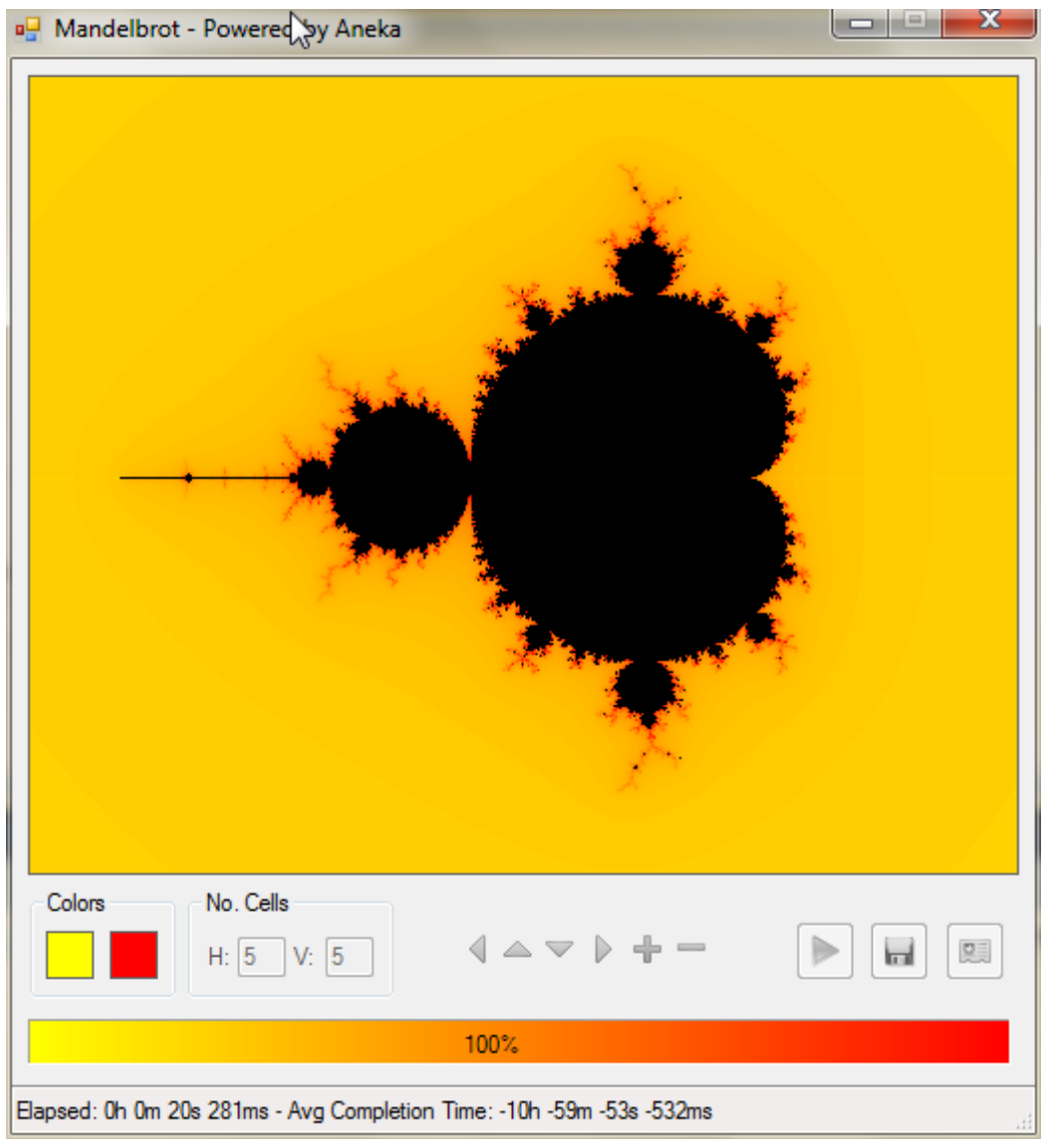

*Figure 20: Mandelbrot Application developing on top of Aneka Thread Model.*

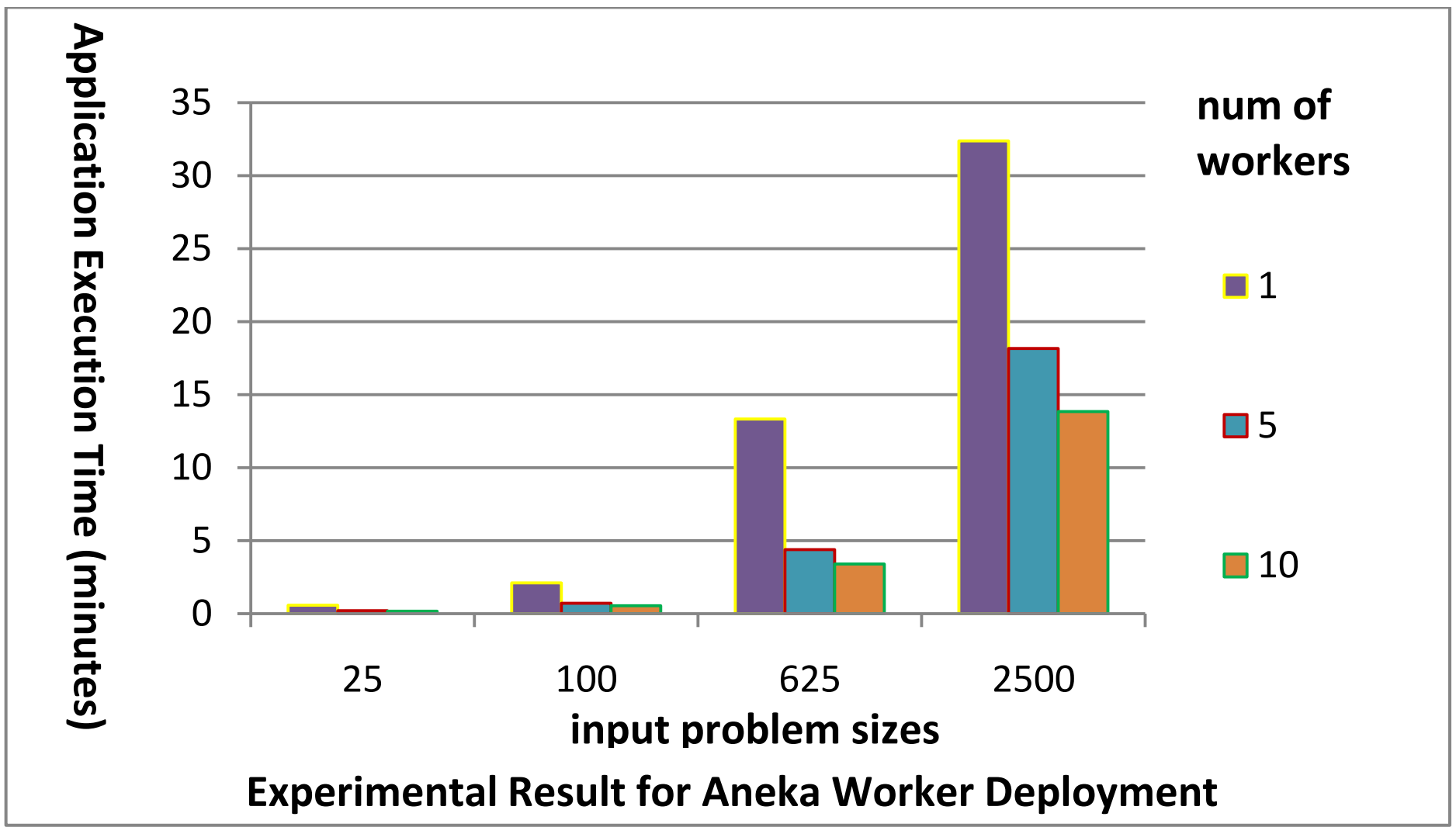

*Figure 21: Experimental Result Showing the Execution Time for Running Mandelbrot Application on Aneka Worker Deployment.*

From both **Figure 21** and **Figure 22**, we can see that for the same input problem size, there is a decrease in the execution time as a result of employing more Aneka Workers to process the work units. The elapsed time used to execute application on Aneka Worker Deployment is also much larger than on Aneka Cloud Deployment due to the communication overhead between the Aneka Master and Aneka Workers with Aneka Workers deployed inside Windows Azure Cloud, while Aneka Master is deployed outside.

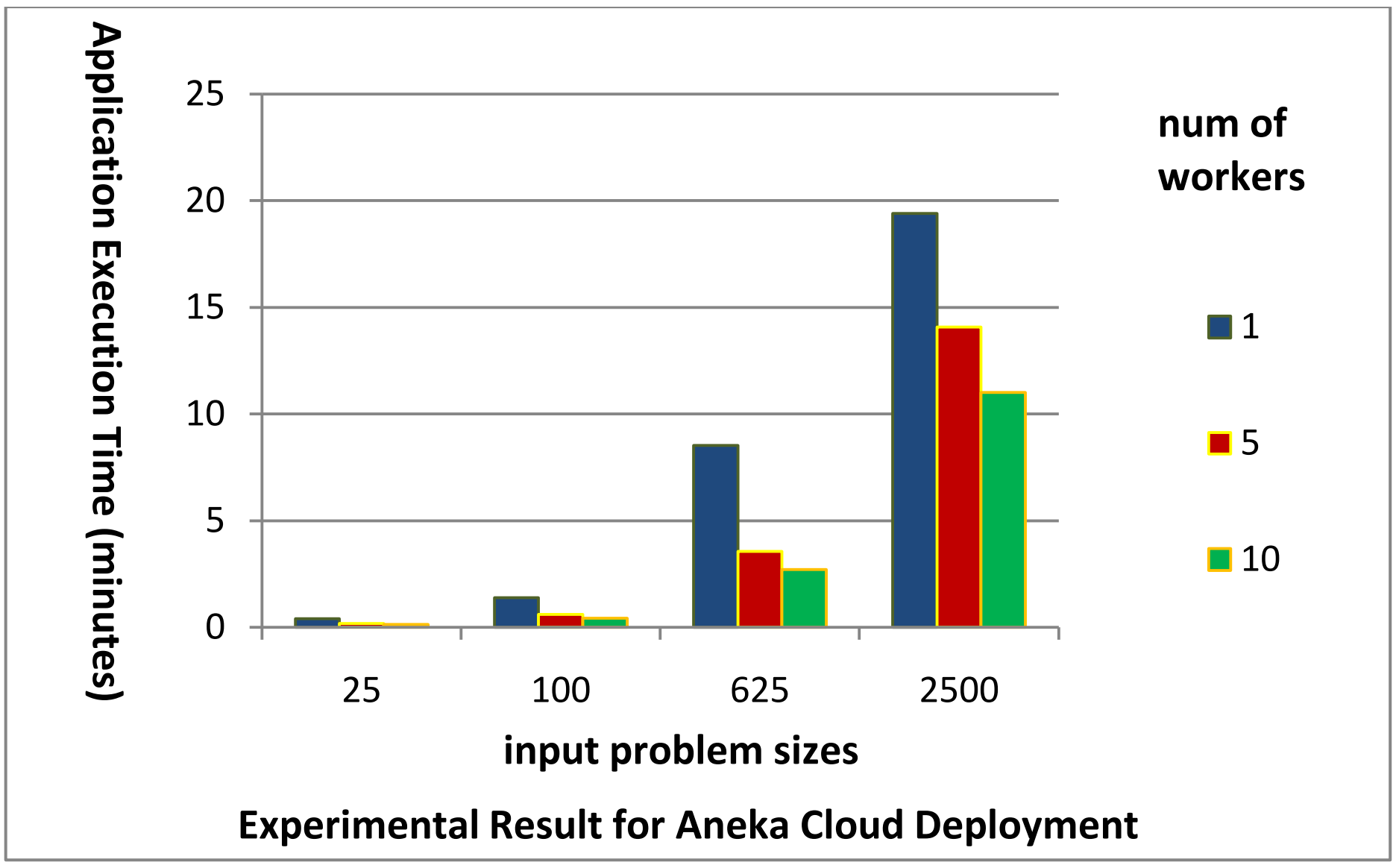

*Figure 22: Experimental Result Showing the Execution Time for Running Mandelbrot Application on Aneka Cloud Deployment.*

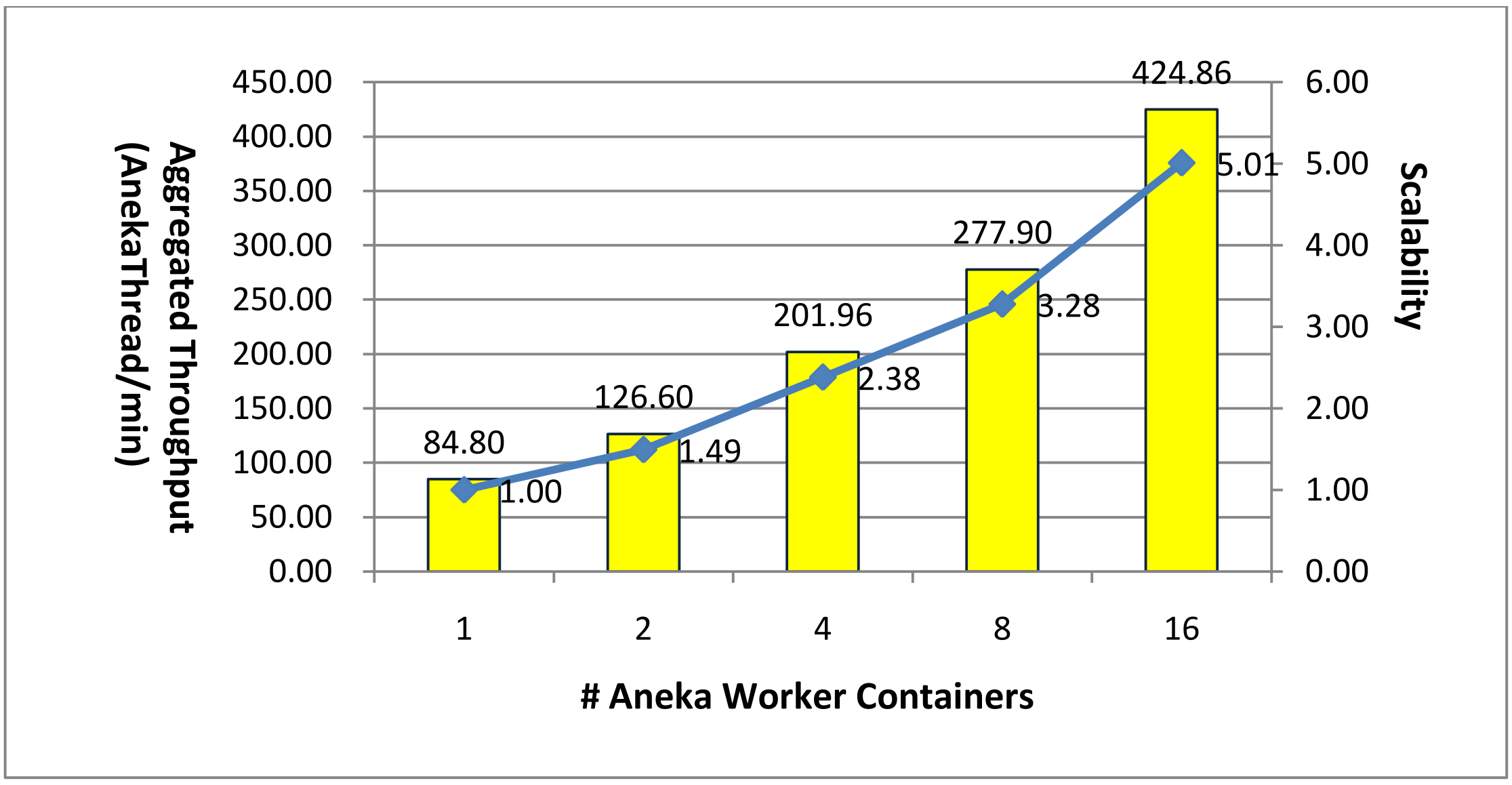

*Figure 23: Scalability Diagram for Aneka Cloud Deployment.*

In the next experiment, we measure the scalability of Aneka Cloud Deployment. In this experiment, we use up to 16 small size instances. All the instances are allocated statically. The result of the experiment is summarized in **Figure 23**. We see that the throughput of the Mandelbrot application running on Azure Cloud Deployment increases when the number of instances ascend.

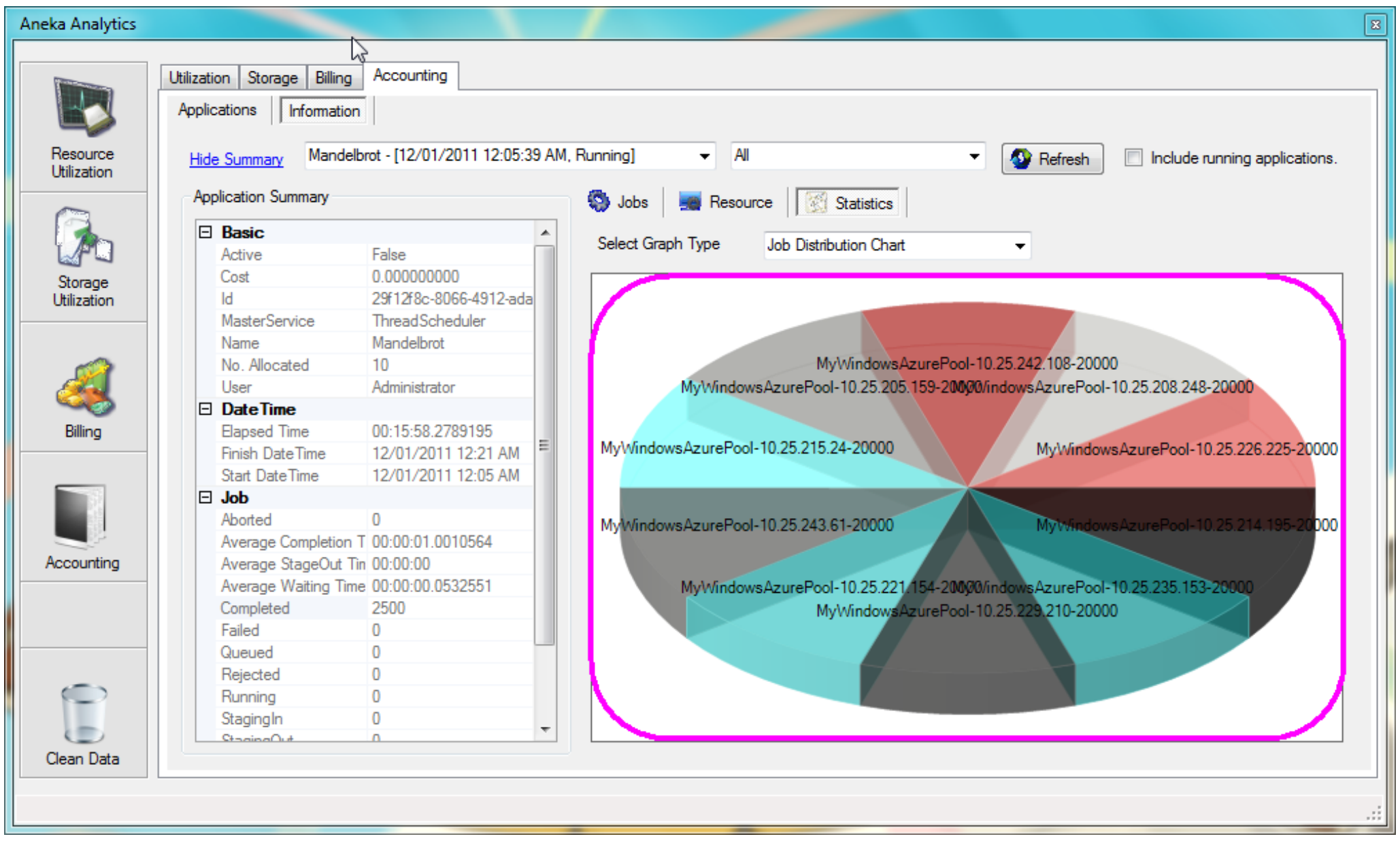

***Figure 24: the Job Distribution Chart shown on the Aneka Analytics Tool***

Furthermore, we can see from ***Figure 24*** that the jobs are evenly distributed across all the available Aneka Workers whose number is 10 in this case.

## 6. RELATED WORK

Windows Azure has been adopted in many projects to build high performance computing applications [7, 8, 9]. Augustyn and Warchał [7] presented an idea and implementation on how to use Windows Azure computing service to solve the N-body problem using Barnes-Hut Algorithm. All computations are operated in parallel on Windows Azure Worker Role instances. Lu et al. [8] delivered a case study of developing AzureBlast, a parallel BLAST engine running on Windows Azure Platform, which can be used to run the BLAST [13], a well-known and both data intensive and computational intensive bioinformatics application. Li et al. [9] demonstrated how to build the MODIS Satellite Data Reprojection and Reduction Pipeline on Windows Azure.

In these cases, the whole implementation is started from scratch, which means the developers need to handle application administration, task scheduling, communication and interaction between role instances, and the storage service access. The Aneka PaaS integration with Windows Azure Platform can speed up the entire development for high performance application running on top of Windows Azure by using the programming models powered by Aneka.

Besides, similar to Aneka, Lokad-Cloud [10] is an execution framework which provides build-in features such task scheduling, queue processing and application administration, and allows users to define a set of services to be run in Windows Azure Worker Role instances. Nevertheless, different from the Aneka PaaS, Lokad-Cloud is only designed to run applications on top of Windows Azure. It is worth mentioning that Aneka PaaS is designed to run applications on private Cloud as well as on public Clouds such as Windows Azure and Amazon EC2. Aneka PaaS can be leveraged to integrate private Clouds with public Clouds by dynamically provisioning resources on public Clouds such as Windows Azure when local resources cannot meet the computing requirement.  Moreover, Aneka supports three types of programming models, the Task Model, Thread Model and MapReduce Model, to meet the requirements of different application domains.

## 7. SAMPLE APPLICATIONS OF ANEKA

Different from other Cloud platforms or Cloud applications running on top of Windows Azure we introduced in **Section** 6, Aneka allows seamless integration of public Clouds and private Clouds to leverage their resources to executing applications. Specifically, a wide range of applications from scientific applications, business services, to entertainment and media, or manufacturing and engineering applications have benefited from Aneka PaaS. A list application types that utilised Aneka is shown in **Table 1**.

## 8. CONCLUSIONS AND FUTURE DIRECTIONS

In this chapter, we have introduced the Aneka Cloud Application Development Platform (Aneka PaaS), presented and discussed the background, design and implementation of the integration of the Aneka PaaS and Windows Azure Platform.

The Aneka PaaS is built on a solid .NET service oriented architecture allowing seamless integration between public Clouds and mainstream applications. The core capabilities of the framework are expressed through its extensible and flexible architecture as well as its powerful application models featuring support for several distributed and parallel programming paradigms. These features enhance the development experience of software developers allowing them to rapidly prototype elastically scalable applications. Applications ranging from the media and entertainment industry, to engineering, education, health and life sciences and several others have been proven to be appropriate to the Aneka PaaS.

Admittedly, the integration of two platforms would give numerous benefits to not only the users of Aneka PaaS but also the customers of Windows Azure Platform, enabling them to embrace the advantages of Cloud computing in terms of more computing resources, easier programming model, and more efficiency on application execution at lower expense and lower administration overhead.

In the first stage, we deployed the Aneka Worker Container as instances of Windows Azure Worker Role, as well as support for dynamic provisioning of Aneka Workers on Windows Azure. In the second step, we deployed the Aneka Master Container on Windows Azure as an instance of Worker Role and the entire Aneka PaaS ran completely on the Windows Azure Platform. This allows users to run Aneka Cloud applications without requiring any local infrastructure. The message transfer overhead and the transfer cost will decrease dramatically. This is beneficial to both Service Providers who uses Aneka PaaS to deliver their services and the final users who consume the services.

**Table 1: Sample Applications of Aneka**

| Industry Sectors | Challenges and Issues | Aneka PaaS Usage |
|---|---|---|
| 1. Geospatial Sciences and Technologies | More geospatial and non-spatial data is involved due to increase in the number of data sources and advancement of data collection methodologies. | • Enable a new approach to complex analyses of massive data and computationally intensive environments.<br>• Build a high-performance and distributed GIS environment over the public, private and hybrid Clouds. |
| 2. Health and Life Sciences | High volume and density of data require for processing. | • Enable faster execution and massive data computation.<br>• Suitable for life science R&D, clinical simulation, and business intelligence tools. |
| 3. Financial Services | Applications such as portfolio and risk analysis, credit fraud detection, option pricing require the use of high-performance computing systems and complex algorithms. | • Simplify the application development lifecycle.<br>• Reduce hardware investment.<br>• Lower ongoing operational expenditure.<br>• Brings a breakthrough in industry standard tools for financial modelling such as Microsoft Office Excel by solving its computational performance barrier. |
| 4. Telecom Industry | The majority of Telecom providers have several disparate systems and they don't have enough capacity to handle the utilization and access information to optimize their use. | • Help telecom providers to realize system utilization strategies in a cost effective, reliable, scalable and tightly integrated manner.<br>• Help mission-critical applications by automating their initiation across a shared pool of computational resources, by breaking the executions into many parallel workloads that produce results faster in accordance with agreed upon SLAs and policies. |
| 5. Manufacturing and Engineering | Manufacturing organizations are faced with a number of computing challenges as they seek to optimize their IT environments, including high infrastructure costs and complexity to poor visibility into capacity and utilization. | • Enable organizations to perform process simulation, modelling, and optimization at a highly increased rate so that the time-to-market of key products is faster, by effectively leveraging Cloud technologies. |
| 6. Entertainment and Media | Business solutions involving digital media transcoding to HD video, 3D image rendering, and gaming, require plenty of time to process and utilize vast amounts of computing capacity to encode and decode the media. | • Optimize networked computers as a private Cloud or leverage public Cloud such as Windows Azure, Amazon EC2 and Go-Grid.<br>• Allows scaling applications linearly.<br>• Better utilize the Cloud farm providing best efficiency and speed possible using Cloud scalability. |

On the whole, in addition to the integration with Windows Azure Platform, presently, Aneka PaaS has already supported the integration of Amazon EC2, GoGrid, and Xen Server. The support of provisioning resources on Windows Azure Platform once again illustrates the adaptability, flexibility, mobility and extensibility of the Aneka PaaS. In the next stage, the Aneka PaaS will continue to integrate with other public Cloud platforms and virtual machine management platforms such as VMWare, Microsoft HyperV and so forth, to help users to exploit more power of Cloud computing.

## ACKNOWLEDGEMENT

This work is partially supported by a grant from the Australian Department of Innovation, Industry, Science and Research via its COMET (Commercialising Emerging Technologies) Program.